\documentclass[
reprint,
superscriptaddress,
amsmath,amssymb,
aps,
pre,
floatfix,
]{revtex4-2}%

\usepackage{graphicx}%
\graphicspath{{figures/}} 
\usepackage{dcolumn}%
\usepackage{bm}%
\usepackage{todonotes}
\usepackage{longtable}
\usepackage{siunitx}
\usepackage{ulem}

\usepackage[colorlinks = true,
linkcolor = blue,
urlcolor  = blue, 
citecolor = blue,
anchorcolor = blue]{hyperref}%

\usepackage{float}

\usepackage{enumitem,amssymb}
\newlist{todolist}{itemize}{2}
\setlist[todolist]{label=$\square$}
\usepackage{pifont}
\begin{document}

\preprint{APS/123-QED}

\title{Efficacy of the Radial Pair Potential Approximation for Molecular Dynamics Simulations of Dense Plasmas}

\author{Lucas J. Stanek}
\email{staneklu@msu.edu}
\affiliation{Department of Computational Mathematics, Science and Engineering, Michigan State University, East Lansing, MI 48824, USA}
\affiliation{Sandia National Laboratories, Albuquerque, NM 87185, USA}

\author{Raymond C. Clay III}
\email{rclay@sandia.gov}
\affiliation{Sandia National Laboratories, Albuquerque, NM 87185, USA}%

\author{M.W.C. Dharma-wardana}
\affiliation{National Research Council of Canada, Ottawa K1A 0R6, Canada}%

\author{Mitchell A. Wood}
\affiliation{Sandia National Laboratories, Albuquerque, NM 87185, USA}%

\author{Kristian R.C. Beckwith}
\affiliation{Sandia National Laboratories, Albuquerque, NM 87185, USA}

\author{Michael S. Murillo}
\email{murillom@msu.edu}
\affiliation{Department of Computational Mathematics, Science and Engineering, Michigan State University, East Lansing, MI 48824, USA}

\begin{abstract}
Macroscopic simulations of dense plasmas rely on detailed microscopic information that 
can be computationally expensive and is difficult to verify experimentally. In this work, we delineate the accuracy boundary between microscale simulation methods by comparing Kohn-Sham density functional theory molecular dynamics (KS-MD) and radial pair potential molecular dynamics (RPP-MD) for a range of elements, temperature, and density. By extracting the optimal RPP from KS-MD data using force-matching, we constrain its functional form and dismiss classes of potentials that assume a constant power law for small interparticle distances. Our results show excellent agreement between RPP-MD and KS-MD for multiple metrics of accuracy at temperatures of only a few electron volts. The use of RPPs offers orders of magnitude decrease in computational cost and indicates that three-body potentials are not required beyond temperatures of a few eV. Due to its efficiency, the validated RPP-MD provides an avenue for reducing errors due to finite-size effects that can be on the order of $\sim20\%$. 
\end{abstract}

\maketitle

\section{\label{sec:intro}Introduction}
High energy-density science relies heavily on computational models to provide information not accessible experimentally due to the high pressure and transient environments. The plasmas in these experiments typically contain strongly coupled ions and partially-degenerate electrons, which constrains our microscopic modeling choices to molecular dynamics (MD) and Monte Carlo approaches. Interfacial mixing in warm dense matter, for example, requires costly, large-scale MD simulations; but, such simulations reveal previously unknown transport mechanisms \cite{stanton2018multiscale}. It is therefore crucial to quantify the efficacy of computational models in different regions of species-temperature-density space so that the cheapest accurate model can be exploited to address such problems. While it is desirable to use short-range, radial, pair potentials (RPPs) to maximize the length and time scales, $N$-body energies may be required in some cases. Few studies have been carried out that comprehensively assess the limitations of RPPs and the regimes of utility for the extant forms; given that the force law is the primary input into MD models, it essential to have quantitative information about these force models.

A wide variety of RPPs have been developed for modeling dense plasmas. In some cases the accuracy of the model can be inferred from its theoretical underpinnings; in other cases, comparison to higher-fidelity approaches or experiments is needed. 
Limitations of the RPP approximation are generally unknown unless compared to an $N$-body potential simulation result such as KS-MD.
Both KS-MD simulations and this comparison are
time-consuming processes that are limited to the temperature regime in which the pseudopotentials necessary for KS-MD are valid. Moreover, comparisons between RPP-MD and KS-MD are limited in the literature, have not been carried out for a range of elements and temperatures, and are often validated with integrated quantities where individual particle dynamics have been averaged and results are subject to cancellation of errors. 

In this work, we carry out KS-MD simulations for a range of elements, temperature, and density, allowing for a systematic comparison of three RPP models. While multiple RPP models can be selected, we choose to compare the widely used Yukawa potential, which accounts for screening by linearly perturbing around a uniform density in the long-wavelength (Thomas-Fermi) limit, a potential constructed from a neutral pseudo-atom (NPA) approach \cite{harbour2016pair,dharma2012electron, perrot1995equation,harbour2018ion}, and the optimal force-matched RPP that is constructed directly from KS-MD simulation data.

With this choice of RPP models, we constrain the functional form of the RPP required to accurately describe microscopic interactions for the systems studied here. For example, if the forced-matched RPP cannot accurately reproduce the metrics described herein, it is very likely that no RPP can describe the system and a KS-MD simulation, or perhaps an interaction potential beyond a RPP (e.g. three-body interaction potentials) is needed. Additionally, if NPA results agree with KS-MD, we are able avoid the force-matching process which requires data from KS-MD. Furthermore, if a Yukawa model agrees with KS-MD, we can neglect including a pseudopotential in the construction of the RPP, reducing the interaction to a simple, analytic form.

Using multiple metrics of comparison between RPP-MD and KS-MD including the relative force error, ion-ion equilibrium radial distribution function $g(r)$, Einstein frequency, power spectrum, and the self-diffusion transport coefficient, the accuracy of each RPP model is analyzed. By simulating disparate elements, namely an alkali metal, multiple transition metals, a halogen, a non-metal, and a noble gas, we see that force-matched RPPs are valid for simulating dense plasmas at temperatures above fractions of an eV and beyond. We find that for all cases except for low temperature carbon, force-matched RPPs accurately describe the results obtained from KS-MD to within a few percent. By contrast, the Yukawa model appears to systematically fail at describing results from KS-MD at low temperatures for the conditions studied here validating the need for alternate models such as force-matching and NPA approaches at these conditions.

In Sec. \ref{sec:meth} we discuss how RPPs arise from second order perturbation theory and how their representation influences the shape of $g(r)$ due to particle crowding and/or attraction. Comparisons between RPPs and KS-MD are done in Sec. \ref{sec:results}, where we begin by comparing interparticle forces illustrating how an increase in temperature indicates an increase in accuracy. In addition the $g(r)$, Einstein frequency, power spectrum, and self-diffusion coefficient are compared, highlighting how an approximately accurate $g(r)$ does not ensure similar accuracy in time correlation functions and transport coefficients. A description of how we accurately compute the self-diffusion coefficient and its uncertainty when finite-size errors are non-negligible is given in Sec. \ref{sec:SD}. This further emphasises the need for RPPs, as we minimize finite-size errors in KS-MD simulations by making the necessary corrections as shown in Sec. \ref{sec:FSE}. We conclude by comparing fully converged (in particle number and simulation time) self-diffusion coefficients to an analytic transport theory; benchmarking its accuracy and providing an effective interaction correction to extend the range of applicability.

\section{\label{sec:meth}Models for the Interaction Potential}

\begin{figure}
    \centering
    \includegraphics[width=8.6 cm]{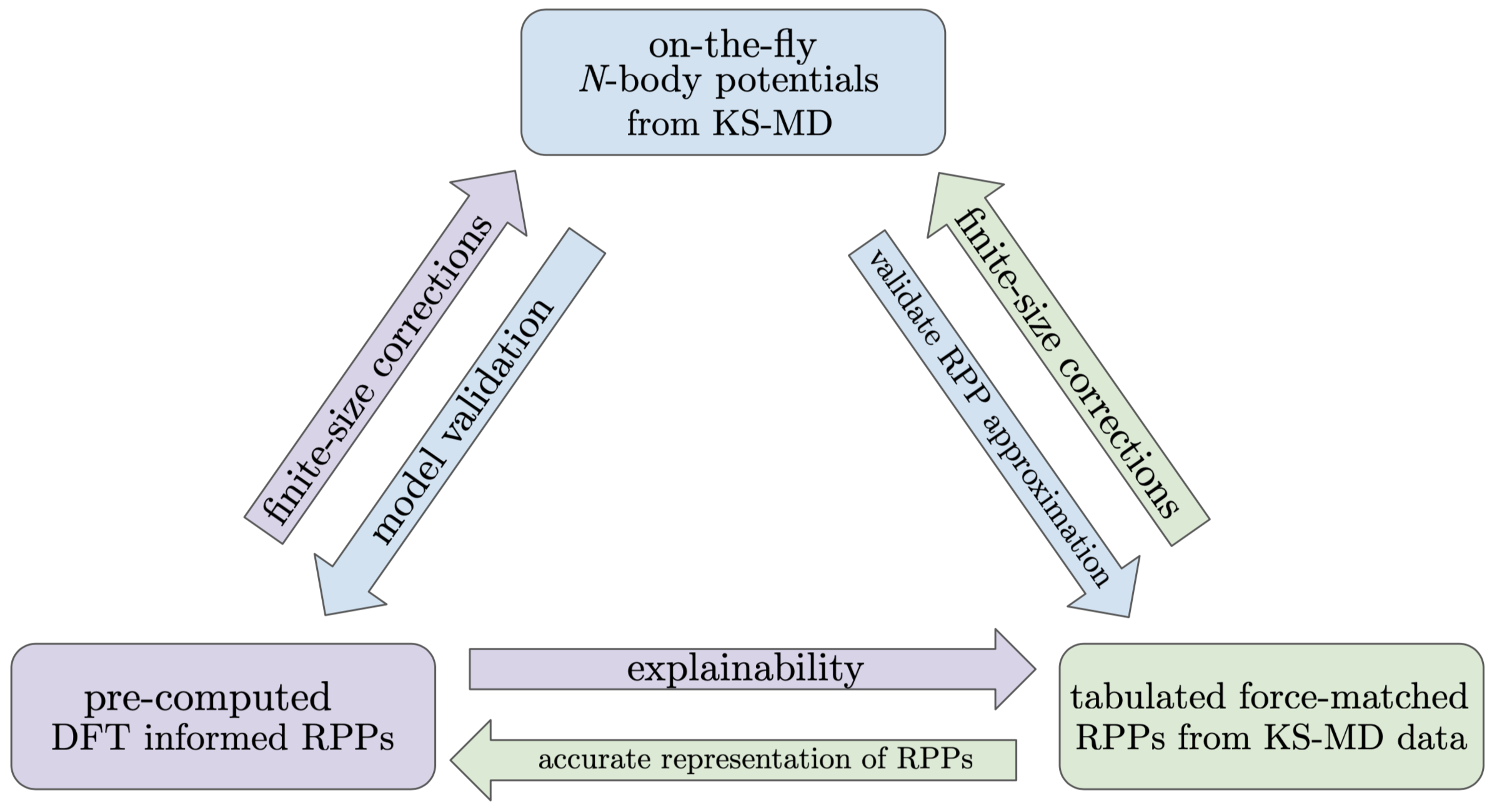}
    \caption{Connections between different portions of this work. $N$-body potentials, shown in the top box, are used to validate pair potential models (lower right) and produce optimal tabulated potentials (lower left).}
    \label{fig:outline}
\end{figure}

The theoretical foundations of the models we will compare are described in this section; their connections are shown in Fig. \ref{fig:outline}. We compare three classes of interactions that are based on the ionic $N$-body energy, shown in the top box, pair interactions that are pre-computed and are analytic or tabulated, shown in the lower-left box, and optimal pair interactions extracted from the $N$-body results, shown in the lower-right box. By comparing these three approaches we aim to answer several specific questions. First, given the nuclear charge $Z$, ionic number density $n_i$, and temperature $T$, what ranges in $\{Z, n_i, T\}$ space are the fast, pre-computed interactions valid and therefore allow for large-scale heterogeneous simulations? Second, how accurate is the ``optimal" pair interaction, and what do its limitations reveal about the need for three-body interactions (and perhaps beyond)? Can these interactions be used to test and correct for finite-size errors? Third, can the optimal interactions guide the development of pre-computed interactions? To simplify the discussion we will consider single species matter with a range of $Z$, each species at its normal solid ionic mass density $\rho_i$, or in some cases half of that, and in thermodynamic equilibrium at temperature $T$. While we do not consider mixtures in this work, the framework is general and can be straightforwardly applied to them. 

Assuming the Born-Oppenheimer approximation holds, we define a potential energy surface for the ions as
\begin{equation}
\label{Nbody}
U_{tot} = U_N(\mathbf{r}_1,\mathbf{r}_2, \ldots, \mathbf{r}_{N}).  
\end{equation}
Physically, the ionic potential energy surface is determined by the electronic charge distribution arising from ions at a particular set of coordinates; in general, (\ref{Nbody}) does not simplify into sums over pairwise terms. There are two major approaches to obtaining (\ref{Nbody}) in practice. The approach represented by the top box in Fig. \ref{fig:outline} computes the electronic charge distribution for each ionic distribution. This is achieved computationally in Kohn-Sham approaches by reducing the electron many-body problem to a single-electron problem in which the Kohn-Sham electron moves in the external field of $N$-ionic centers. The dominant computational cost comes from solving an $N_o \times N_o$ set of eigenvalue equations, where $N_o$ is the number of single-particle orbitals. Even though the electron many-body problem has been simplified to a one-body problem, matrix diagonalization incurs a cost of $O(N_o^3)$, and at high temperatures the smearing of the Fermi-Dirac distribution requires an increasing  number of orbitals leading to significant increases in computational cost. The complexity of the electron charge distribution also demands the use of an advanced ``Jacob's ladder" of exchange-correlation functions to address the electron many body problem. 

This approach yields an intrinsically ionic $N$-body potential energy surface; the electronic density is computed using a description appropriate to the choice of $\{Z,n_i,T\}$. The second approach to calculating the potential energy surface is to use a cluster-type expansion, which takes the form
\begin{equation}
\label{n_bodies}
	U_{tot} = \sum_{i}^NU_1(\mathbf{r}_i) + \sum_{i, j}^N U_2(\mathbf{r}_i, \mathbf{r}_j) + \sum_{i, j, k}^N U_3(\mathbf{r}_i, \mathbf{r}_j, \mathbf{r}_k) +  \cdots.
\end{equation}
When this expansion can be truncated with only a few terms, interactions can be pre-computed and fast neighbor algorithms allow for a very rapid evaluation of forces, typically many orders of magnitude faster than through use of (\ref{Nbody}). This allows, for example, for simulations with trillions of particles \cite{germann2008trillion,eckhardt2013591,heinecke2015supercomputing}. However, the disadvantages are that the computational cost increases rapidly as more terms are included and the accuracy of a specific truncation and choice of functional forms with that truncation are not usually known; part of our goal is to assess how accurate the potential energy surface in (\ref{Nbody}) can be represented by the first two terms of (\ref{n_bodies}).

\subsection{\label{sec:nbody}{\it{N}}-body Interaction Potentials}

The most accurate forces are obtained from the gradient of the total energy in (\ref{Nbody}), which requires the entire ionic configuration. Although machine learning approaches are enabling the ability to pre-learn that relationship \cite{glielmo2018efficient,chmiela2018towards,schmidt2019recent}, it remains more common to compute the forces for each ionic configuration during the simulation (``on-the-fly"). We obtain the electronic number density for each ionic configuration in the Kohn-Sham-Mermin formulation of the density
\begin{align}
    n_e({\bf r}) = \sum_i f_i(T)|\phi_i({\bf r})|^2,
\end{align}
where $T$ is the temperature of the system in energy units, the Fermi occupations are given by $f_i(T) = (1 + e^{\beta(E_i - \mu)})^{-1}$, and the Kohn-Sham-Mermin orbitals $\phi_i({\bf{r}})$ satisfy
\begin{equation}\label{eq:kohn-sham-eq}
\left( -\frac{1}{2}\nabla^2 + v_{eff}(\mathbf{r}) \right) \phi_i(\mathbf{r}) = \epsilon_i \phi_i(\mathbf{r}),
\end{equation}
where 
\begin{equation}\label{eq:effective_potential}
v_{eff}(\mathbf{r}) = V_{ext}(\mathbf{r}) + \int\! d\mathbf{r}'\:\left[ \frac{n_e(\mathbf{r}')}{ |\mathbf{r}-\mathbf{r}'|} + \frac{\delta E_{xc}[\rho]}{\delta \rho(\mathbf{r})}\right],
\end{equation}
is a sum of the external ($N$ ion-electron), Hartree, and exchange-correlation energies. Our KS-MD simulations were done using the Vienna Ab-initio Simulation Package (VASP) \cite{PhysRevB.47.558, KRESSE199615, PhysRevB.54.11169, PhysRevB.59.1758}. The finite temperature electronic structure was treated with Mermin free-energy functional, and we used the Perdew–Burke-Ernzerhof functional for the exchange correlation energy \cite{PhysRevLett.77.3865}.  To improve computational efficiency, we eliminated the chemically inactive core electrons with projector augmented-wave \cite{PhysRevB.50.17953} pseudopotential, the details of which are given in the supplemental information \cite{PhysRev.139.A796}. Sixty-four atoms ($N=64$) were used in these simulations, with an energy cutoff of $800$ eV and at the Baldereschi mean-value k-point \cite{PhysRevB.7.5212} for all temperatures ranging from $T = 0.5$ to $15$ eV. A simulation time step of $0.1$ fs was used and the total simulation lengths for each case vary and are on the order of a few picoseconds. All KS-MD simulations were first equilibrated in the NVT ensemble and then carried out in the NVE ensemble where data was collected.

\subsection{Force-Matching}
After the Kohn-Sham potential energy surface has been computed, we aim to construct a compact representation of (\ref{Nbody}) with (\ref{n_bodies}). By assuming an approximate functional form for (\ref{n_bodies}) and defining a suitable cost-function based on the difference between the Kohn-Sham and approximate model forces, we use the force matching procedure \cite{ercolessi1994interatomic,  doi:10.1063/1.1739396, doi:10.1080/14786430500333349, brommer2007potfit, Brommer_2015} to generate the optimal RPP model based on the KS-MD force data. By choosing an optimizable spline form for the RPP the explicit form of the model is entirely determined from the KS-MD force data and not limited to a fixed functional form.

While the force-matched RPP yields the best RPP to reproduce the KS-MD force data, it could be the case 3-body and higher interactions are non-negligible. To check this, we selectively employ the Spectral Neighborhood Analysis Potential (SNAP) which constructs a potential energy surface from a set of four-body descriptors (bispectrum components) where each descriptor is independently weighted and these weights are determined by regressing against KS-MD data of energies and forces. A descriptor captures the strength of density correlations between neighboring atoms and the central atom within a given cutoff distance, details can be found in \cite{THOMPSON2015316, wood2018extending}. SNAP potentials utilizing $56$ bispectrum component descriptors were trained on $10\%$ of the KS-MD dataset, and additionally tested against an additional $10\%$ to ensure regression errors were properly minimized and avoided over-fitting of the KS-MD data. %

\subsection{\label{sec:pairpots}  Radial Pair Potentials}

As the computational cost of using on-the-fly $N$-body interactions is often prohibitive, the least expensive approach utilizes pre-computed RPPs ignoring most of the terms in (\ref{n_bodies}). Many functional forms for the RPP have been proposed for application to warm dense matter often using the second-order perturbation-theory interaction energy
\begin{align}
\label{sopt}
    u(k) &= \langle Z\rangle^2 u_C(k) + \left|u_{ei}(k) \right|^2 \chi(k),
\end{align}
which is the standard Fourier-space result \cite{porter2010pair} written in terms of the mean ionization state $\langle Z \rangle$, the bare Coulomb potential $u_C(k) = 4\pi e^2/k^2$, the electron-ion pseudopotential $u_{ei}(k)$ and the susceptibility $\chi(k)$. 

In practice, pair interactions are constructed using the nearly the same steps as for the $N$-body interactions, with the primary difference being that each ion is replaced with a single ``average atom" (AA), which is an all-electron, non-linear, finite-temperature density functional theory calculation \cite{AA_MIS_PhysRevE.87.063113}; such calculations can also be relativistic \cite{faussurier2019relativistic,sterne2007equation}. 
From the AA, a pseudopotential $u_{ei}(k)$ and an accurate free/valence electron response function $\chi(k)$ are constructed and (\ref{sopt}) is formed. This approach has three strengths: (1) typical AA models are not limited to low temperatures, (2) the interaction (\ref{sopt}) can be pre-computed 
for use in MD,
and (3) pair interactions with a fast nearest neighbor algorithm are very computationally efficient. As we alluded to above, the accuracy loss attendant to these strengths is what we wish to determine in this work. The AA itself is aware of the ionic number density  $n_i$, which sets the ion-sphere radius $a_i = (3/4\pi n_i)^{1/3} $, and includes the fact that there is only one ion in the ion sphere, which implies a $g(r)$; this indirect inclusion of higher-order terms in (\ref{n_bodies}) is true for all AA-based interactions.

Among the simplest variants of (\ref{sopt}) one approximates the pseudopotential as $u_{ei}(k) \approx -4\pi \langle Z\rangle e^2/k^2$, where the mean ionization state $\langle Z \rangle$ results from a AA calculation \cite{murillo2013partial}, and $\chi(k)$ in its long-wavelength (Thomas-Fermi) limit $\chi_{TF}(k)$; this is known as the ``Yukawa" interaction \cite{murillo2000viscosity,stanton2015unified}. Here, we employ a Yukawa interaction with inputs from a Thomas-Fermi AA \cite{stanton2018multiscale}, which we will refer to as ``TFY".  This procedure yields an analytic potential in real space of the form
\begin{equation}
    \label{eq:Yukawa}
    u^{TFY}(r) = \frac{\langle Z\rangle^2 e^2}{r}\exp\left(-r/\lambda_{TF} \right),
\end{equation}
where
\begin{equation}
    \lambda^2_{TF} = \frac{\sqrt{T^2 + (\frac{2}{3}E_F)^2}}{4\pi n_e e^2}, \label{eq:TF_screening}
\end{equation}
and the Fermi energy $E_F = \hbar^2 (3\pi^2n_e)^{2/3}/2m_e$. Note that the TFY interaction is monotonically decreasing (purely repulsive). Computationally, the TFY model is highly desirable because of its radial, pair, analytic form with an exponentially-damped short range. Its weaknesses are the relatively approximate treatments of $u_{ei}(k)$ and $\chi(k)$.  The TFY model can be extended by including the gradient corrections to $\chi_{TF}(k)$, but otherwise retaining the other approximations. This improvement yields the Stanton-Murillo potential \cite{stanton2015unified}; the gradient correction to $\chi_{TF}(k)$ introduces oscillations in the potential in some plasma regimes that are absent in the monotonic TFY model. Moreover, gradient corrections add improvements to the cusp at the origin and the large-$r$ asymptotic behavior. Here, however, we will only employ the simpler TFY model.

A great deal of accuracy can be gained by abandoning analytic inputs to (\ref{sopt}). In this case, self-consistent numerical calculations of each of the terms can be carried out, still allowing for pre-computed interactions; there is essentially no computational overhead for tabulated interactions \cite{wolff1999tabulated}. Here, we employ a NPA model that yields both the mean ionization state and its pseudopotential using a Kohn-Sham-Mermin approach, as described above, but with a finite-temperature exchange-correlation potential; the susceptibility is Lindhard with local field corrections \cite{dharma2012electron}. Note that the electron-ion pseudopotential $u_{ei}(k)$ introduces additional oscillations on length scales different from $\chi(k)$, although the Friedel oscillations in $\chi(k)$ contribute much more to the pair interaction. Note that the name ``NPA" has been used by many authors to several different average-atom models, and many of them involve approximations that limit those models to higher temperatures, e.g., $T>E_F$; however, here we use the one-center density functional theory model developed by Dharma-wardana and Perrot as this model has been tested at high temperatures as well as at very low temperatures, and found to agree closely with more detailed $N$-center density functional theory simulations and path-integral quantum calculations where available.

 \begin{figure}
    \centering
    \includegraphics[width=8.6 cm]{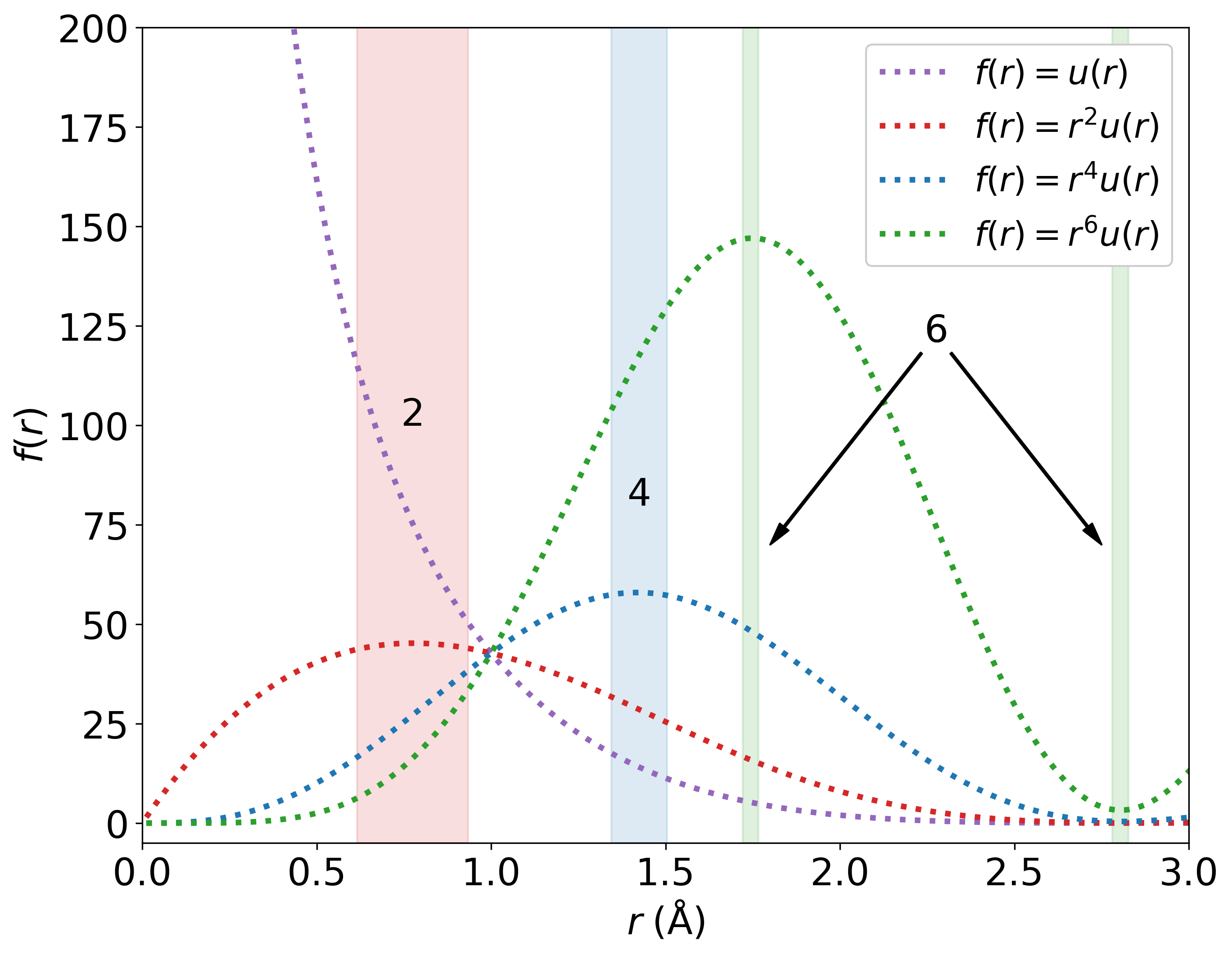}
    \caption{NPA RPP for Al at 2.7 g/cm$^3$ and $T = 1$ eV. Various power laws are valid at different values of $r$. The appropriate power law for a given range of $r$ is shaded and denoted with a ``2", ``4", or ``6".}
    \label{fig:pot_powers}
\end{figure}

\begin{figure}
\centering
\includegraphics[width=8.6 cm]{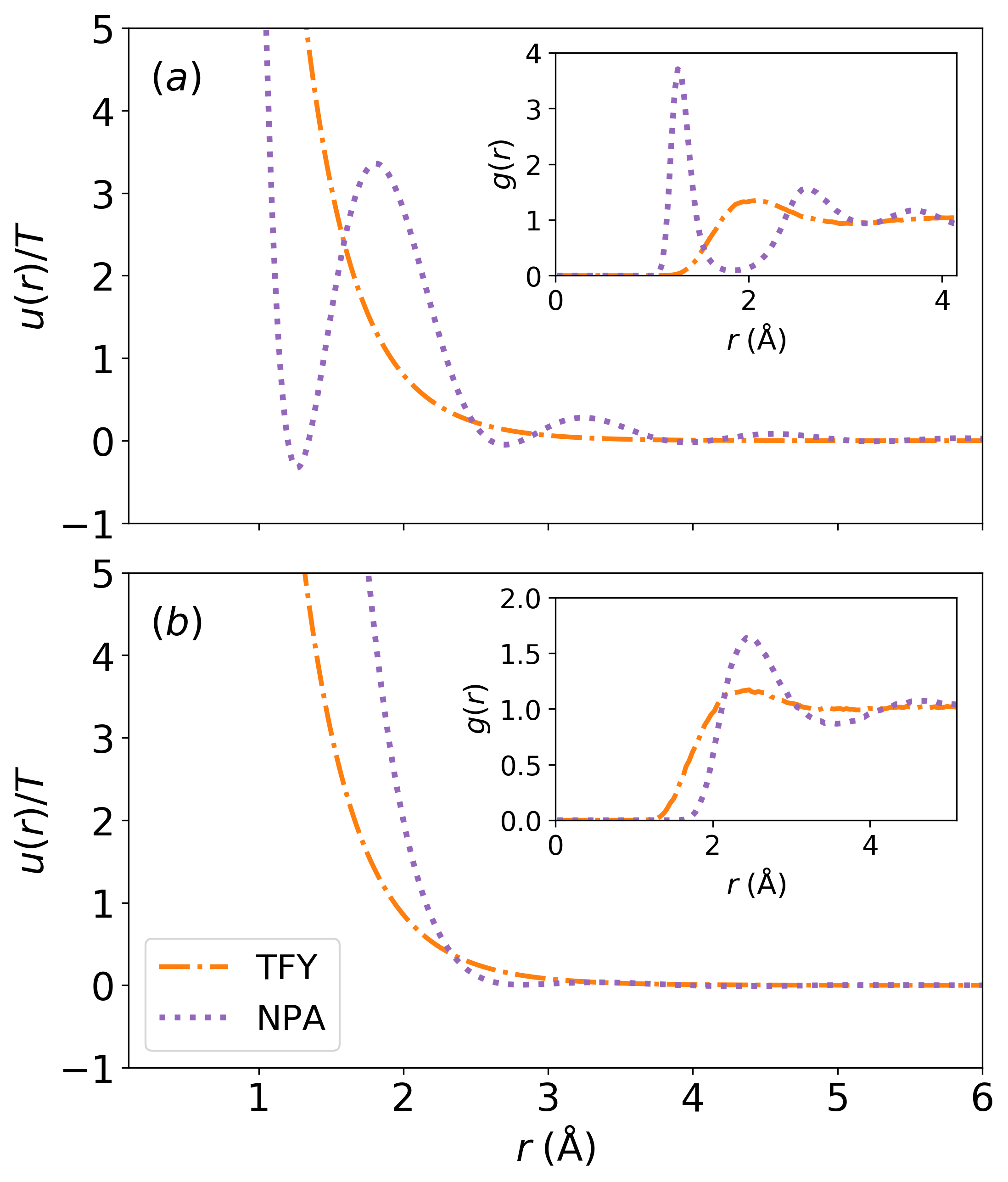}
\caption{Comparison of TFY and NPA RPPs for C at 2.267 g/cm$^3$ and Al at 2.7 g/cm$^3$ with corresponding $g(r)$ computed from MD simulation: $(a)$ $T = 0.5$ eV. The increase in magnitude of the first $g(r)$ peak results from particle attraction. $(b)$ $T = 1$ eV. Particle crowding increases the magnitude of the first $g(r)$ peak.}
\label{fig:CAL_compare}
\end{figure}

It is worth comparing predictions based on (\ref{sopt}) with other forms suggested previously. A popular RPP for warm dense matter studies is the short-range repulsion interaction, which adds a long-range, power-law correction to the TFY model of the form $A/r^4$ \cite{wunsch2009ion, srr_nature, PhysRevLett.110.065001, Glenzer_2016, ruter2014ab, PhysRevLett.103.245004, doi:10.1016/j.mre.2017.09.001}; for $A>0$ this is also a monotonic interaction, with the goal of increasing the strength of the TFY model, which underestimates the peak height of $g(r)$. In Fig. \ref{fig:pot_powers} we examine this ansatz by computing a NPA interaction for Al at solid density and $T = 1$ eV. To find the ``best" power law, we multiply the NPA interaction by various powers $r^a$ to find regions where the interaction is flat; a flat region with $a = 4$ would recover the short-range repulsion interaction. It is clear that the $A/r^4$ is only valid over a very small range of $r$ values; importantly, the NPA interaction shows that $a$ increases with $r$, which is a true short-ranged interaction - the empirical correction the short-range repulsion model adds greatly overestimates the strength of the interaction at large interparticle separations \cite{harbour2016pair}. Worse, the short-range repulsion model potentially gets an accurate answer for the wrong reason, as we explore in Fig. \ref{fig:CAL_compare}. 

\begin{figure*}
    \centering
    \includegraphics[scale=0.35]{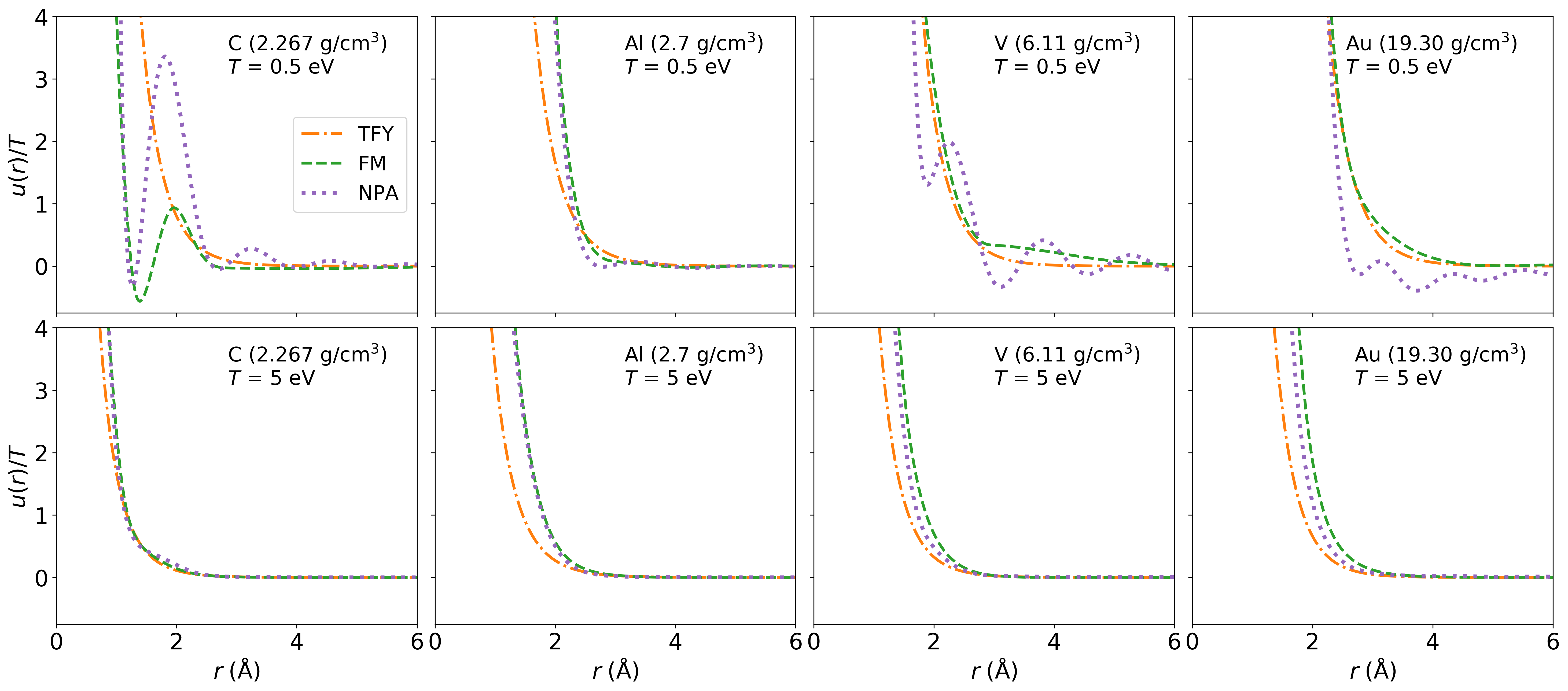}
    \caption{The RPP models normalized by temperature versus distance for C, Al, V, and Au. Top row, $T = 0.5$ eV: the representation of the RPP is element dependent with strong agreement for aluminum. Bottom row, $T = 5$ eV: The agreement between models improves significantly. The differences in the representation can be connected back to (\ref{sopt}) where the treatment of the mean ionization, electron-ion pseudo potential, and susceptibility define the RPP.} 
    \label{fig:ur_compare}
\end{figure*}

Because the form (\ref{sopt}) generally has oscillations, the enhanced peak height of $g(r)$ from the NPA model over the TFY model occurs for two, independent reasons. Attractive regions of the interaction, as shown in the top panel of Fig. \ref{fig:CAL_compare}, can produce very strong peaks in $g(r)$. Conversely, stronger overall repulsion at intermediate $r$ can lead to a similar $g(r)$ behavior, as shown in the bottom panel of Fig. \ref{fig:CAL_compare}, but with rapid decay of the interaction at larger $r$. The functional form (\ref{sopt}) naturally contains both the ``crowding" and ``attraction" behaviors as special cases. Fig. \ref{fig:ur_compare} shows a comparison of the RPPs for C, Al, V, and Au at $T = 0.5$ and $5$ eV. The TFY model is purely monotonic whereas the force-matched and NPA RPPs  attractive have attractive oscillations. Below, we will explore the consequences of these features of the interaction on ionic transport.

Once the RPPs have been constructed, MD simulations were carried out using in the Large-scale Atomic/Molecular Massively Parallel Simulator (LAMMPS) \cite{PLIMPTON19951}. For the tabulated RPPs (force-matched and NPA) a linear interpolation was needed to determine the force value between tabulation points. To make a direct comparison between the RPP-MD and KS-MD results, all simulations were carried out in a 3 dimensional periodic box with 64 atoms and a time step of 0.1 fs. The length of each simulation is identical to the corresponding simulation performed with KS-MD. Keeping these conditions identical avoids the unintentional reduction in statistical errors between KS-MD and RPP-MD. All simulations were first equilibrated in the NVT ensemble so that the average temperature for each simulation during the data collection phase is within 1\% of the reported temperature in Table \ref{tab:transport}. The data collection phase was carried out in the NVE ensemble. In Sec. \ref{sec:FSE} a finite-size effect study was done for the cases of C at 2.267 g/cm$^3$ and V at 6.11 g/cm$^3$ where the total simulation length was increased by 10 times and the number of atoms $N$ increases from $64$ to $256$, $3375$, and $8000$.

\section{\label{sec:results}Results}

\subsection{Force Error Analysis}

\begin{figure*}
    \centering
    \includegraphics[width=17.2 cm]{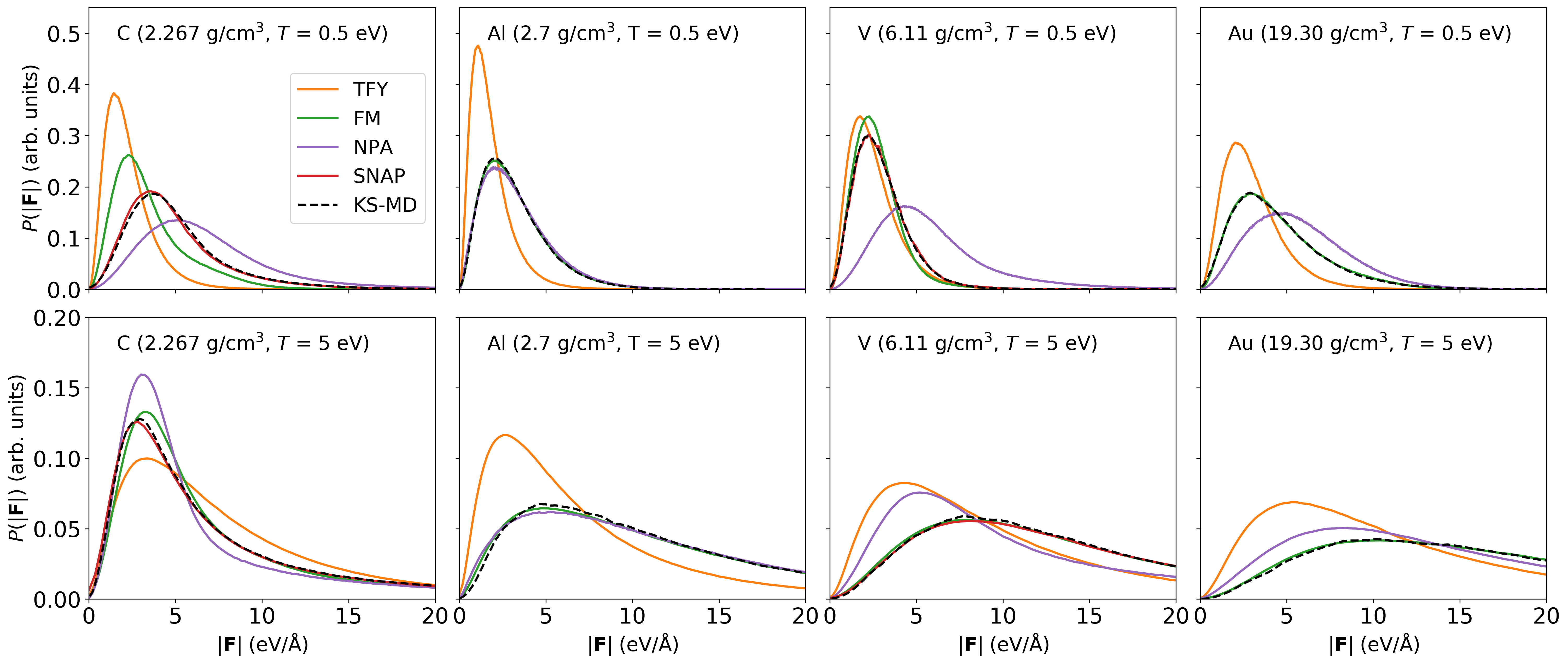} 
    \caption{Microfield distributions for C, Al, V, and Au. The observed trends of the microfields agree with the trends in the trends between the self-diffusion coefficients in Table \ref{tab:transport}. In general, when the microfields are similar to that of KS-MD, the agreement between the self-diffusion coefficient increases. To assess the importance of 3-body or higher interactions, SNAP results are reported for C and V at both $T = 0.5$ and $5$ eV.}
    \label{fig:Pf_compare}
\end{figure*}

\begin{figure*}
    \centering
    \includegraphics[width=17.2 cm]{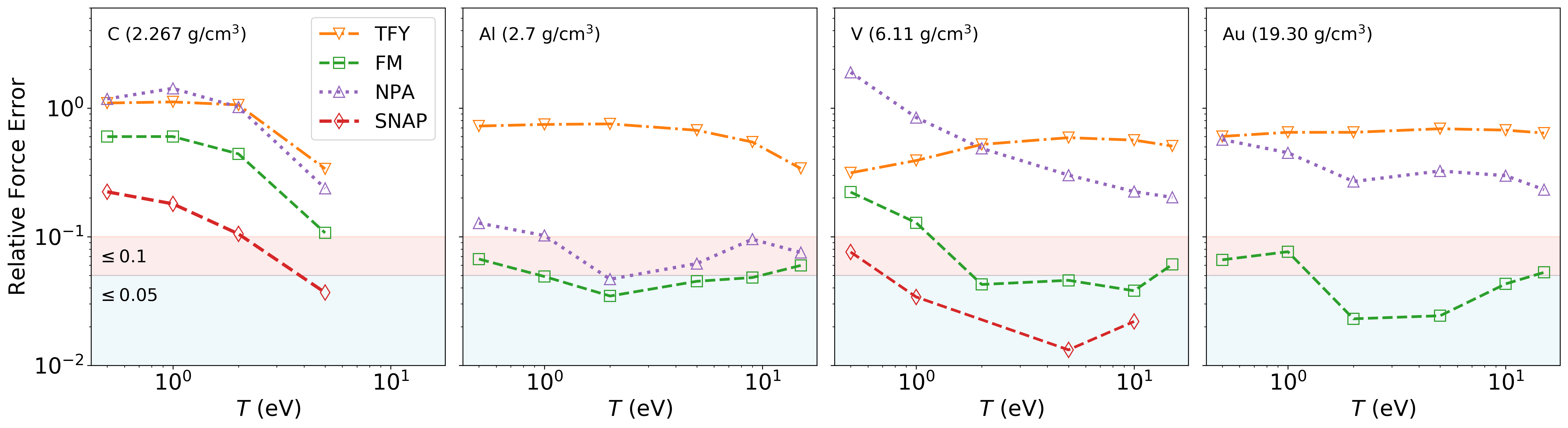}
    \caption{Relative force error versus temperature computed from (\ref{eq:RFE}) for C, Al, V, and Au. The red shaded region indicates force accuracy of $\leq0.1$ and the blue shaded region indicates force accuracy of $\leq 0.05$. SNAP and force-matched RPP yields the lowest relative force error and decreases or remains constant as temperature increases. This indicates an increases in accuracy of the RPP models as temperature increases.}
    \label{fig:f_err}
\end{figure*}

One metric for establishing the accuracy of approximations to the Kohn-Sham potential energy surface is to compute relative force errors between Kohn-Sham force data and an approximate model (RPP or many-body potential) for $M$ particle coordinate configurations. For this, we compute the mean-absolute force error 
\begin{equation}
   MAE = \frac{1}{3MN} \sum_{\alpha,i,m} |F^{(APR)}_{\alpha,i, m}-F^{(KS-MD)}_{\alpha,i,m}|,
   \label{eq:MAE}
\end{equation}
where $F^{(APR)}_{\alpha,i,m}$ and $F^{(KS-MD)}_{\alpha,i,m}$ are the $\alpha$-th force components ($x$, $y$, or $z$) on the $i$-th atom in particle coordinate configuration number $m$ for the approximate model and the KS-MD force data respectively.  

Note that a direct comparison of the mean absolute error between different elements, temperatures, and densities cannot be done as the distribution of forces associated with systems of different elements at different thermodynamic conditions are in general quite different. This can be observed in Fig. \ref{fig:Pf_compare} where a microfield distribution of the force magnitudes is shown. In all cases but C at 2.267 g/cm$^3$ and $T = 5$ eV, the TFY model predicts more frequent small force magnitudes relative to force magnitudes computed from the KS-MD force data. In contrast, for C, V, and Au at $T = 0.5$ eV the NPA RPP tends to predict more large force magnitudes relative to the KS-MD force data. These trends can be connected back to (\ref{sopt}) where the choice of $\langle Z \rangle$, $u_{ei}(k)$, and $\chi(k)$ all contribute to the construction of a RPP model and hence the force magnitudes. More work needs to be done to determine how each term influences the RPP model, the predicted forces, and observables. 

As the microfield force distributions vary for different elements and temperatures, the mean absolute error will also vary. To this end, we seek a scale factor for (\ref{eq:MAE}) to normalize the results across the different elements, temperatures, and densities studied here. Such a scale factor is the ``mean absolute force" defined as

\begin{equation}
    MAF = \frac{1}{3 M N} \sum_{\alpha,i,m} |F^{(KS-MD)}_{\alpha,i,m}| \label{eq:MAF}.
\end{equation}

Using (\ref{eq:MAE}) and (\ref{eq:MAF}) we define the relative force error as
\begin{equation}
RFE =\frac{\sum_{i,\alpha,m} |F^{(APR)}_{\alpha,i,m}-F^{(KS-MD)}_{\alpha,i,m}|}{\sum_{\alpha,i,m} |F^{(KS-MD)}_{\alpha,i,m}|}.\label{eq:RFE}
\end{equation}

This metric has the following desirable property: if the mean absolute error changes with density or temperature in the same way as the underlying force distribution, the relative force error will maintain roughly the same value. Therefore, as we change the thermodynamic conditions for a given element, (\ref{eq:RFE}) provides a temperature independent metric as measured with respect to a KS-MD force data ``baseline". Intuitively, when (\ref{eq:RFE}) evaluates to 1, the mean absolute error is the same order of magnitude as the mean absolute force and when $(\ref{eq:RFE})$ is 0, the approximate model is exactly reproducing the per-component KS-MD force data.    

Fig. \ref{fig:f_err} displays (\ref{eq:RFE}) as a function of temperature for C, Al, V, and Au where general trends can be observed. One trend is that for most RPPs, the relative force error decrease towards higher temperatures, which confirms an intuition long held for the validity of the NPA and TFY models. However, for all systems pictured except C, force-matching drastically reduces the relative force error compared to the NPA and TFY results. Specifically, the force-matched RPPs routinely achieve a relative force error of roughly 0.05 above $T = 5$ eV. Except for the case of the NPA RPP for Al, the NPA and TFY RPPs maintain an error of around 0.2 across the entire the temperature range.

The second major observation from Fig. \ref{fig:f_err} is that while force-matched RPPs drastically lower the observed relative force errors across temperatures compared against other RPPs, we immediately see where a RPP approximation is likely invalid. For example, the relative force error for C using the force-matched RPP is uncharacteristically high (roughly 0.6) until $T = 5$ eV. A similar situation appears for the case of V at $T = 0.5$ eV where the relative force error for the force-matched RPP is roughly 0.25. We can demonstrate explicitly that these discrepancies come from the neglect of 3-body and higher interactions by including relative force errors using a SNAP model. We observe that the relative force error decreases significantly relative to the force-matched RPP. For C, the relative force error drops from roughly 0.6 using a force-matched RPP to 0.2 using a SNAP model at $T = 0.5$ eV. Likewise for V, the relative force error drops from roughly 0.25 using a force-matched RPP to 0.07 using a SNAP model at the same temperature.

\begin{figure}
    \centering
    \includegraphics[width=8.6 cm]{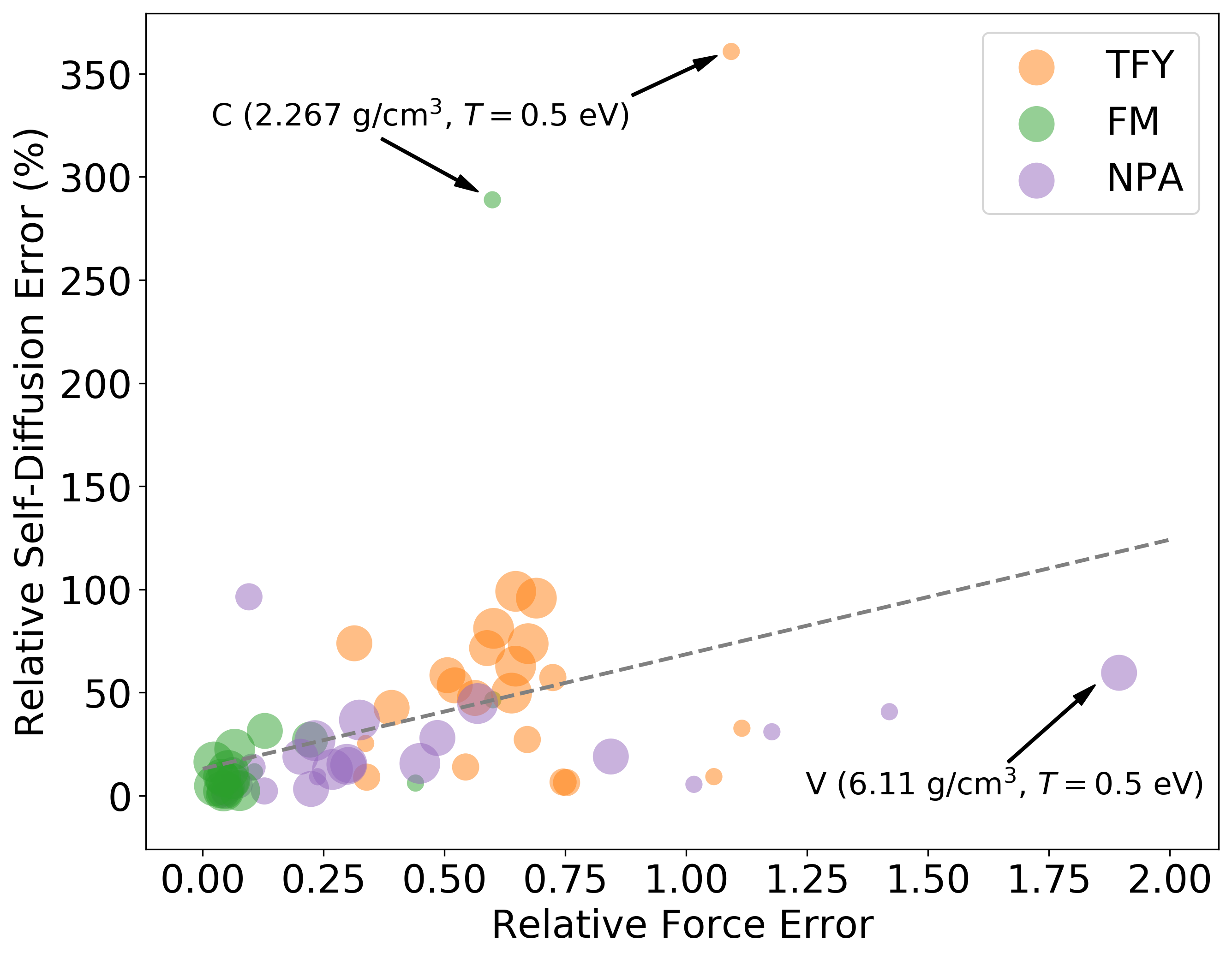}
    \caption{Relative (to KS-MD) self-diffusion error versus the relative force error for C, Al, V, and Au. The size of each point corresponds to the atomic number. The grey dashed line is a linear fit to the points showing a positive correlation between self-diffusion error and force error.}
    \label{fig:SDf_err}
\end{figure}

Ultimately, it is not the component-wise force or the interaction potential we care about generating, but rather observables such as $g(r)$ and self-diffusion coefficient. To address this connection, we examine correlations between the force error and the self diffusion coefficient error, as shown in Fig. \ref{fig:SDf_err}. While there is a general trend with increasing errors in both quantities (shown with a linear fit), there are also some clear outliers. For the case of C at 2.267 g/cm$^3$ and $T = 0.5$ eV, we find that the NPA and TFY RPPs, which reproduce the KS-MD reference forces to within a comparable amount relative to other elements, produce a self-diffusion coefficient that differs from the KS-MD result by many factors. This case is marked with arrows in Fig. \ref{fig:SDf_err}. Conversely, for V at $T = 1$ eV, the relative self-diffusion percent error is low, yet the relative force error is high. The imperfect mapping of relative self diffusion error versus relative force error suggests that physics beyond a RPP is needed, possibly at least a three-body angular dependence, but further work is needed. 

\subsection{Radial Distribution Function and The Einstein Frequency}

\begin{figure*}
    \centering
    \includegraphics[width=17.2 cm]{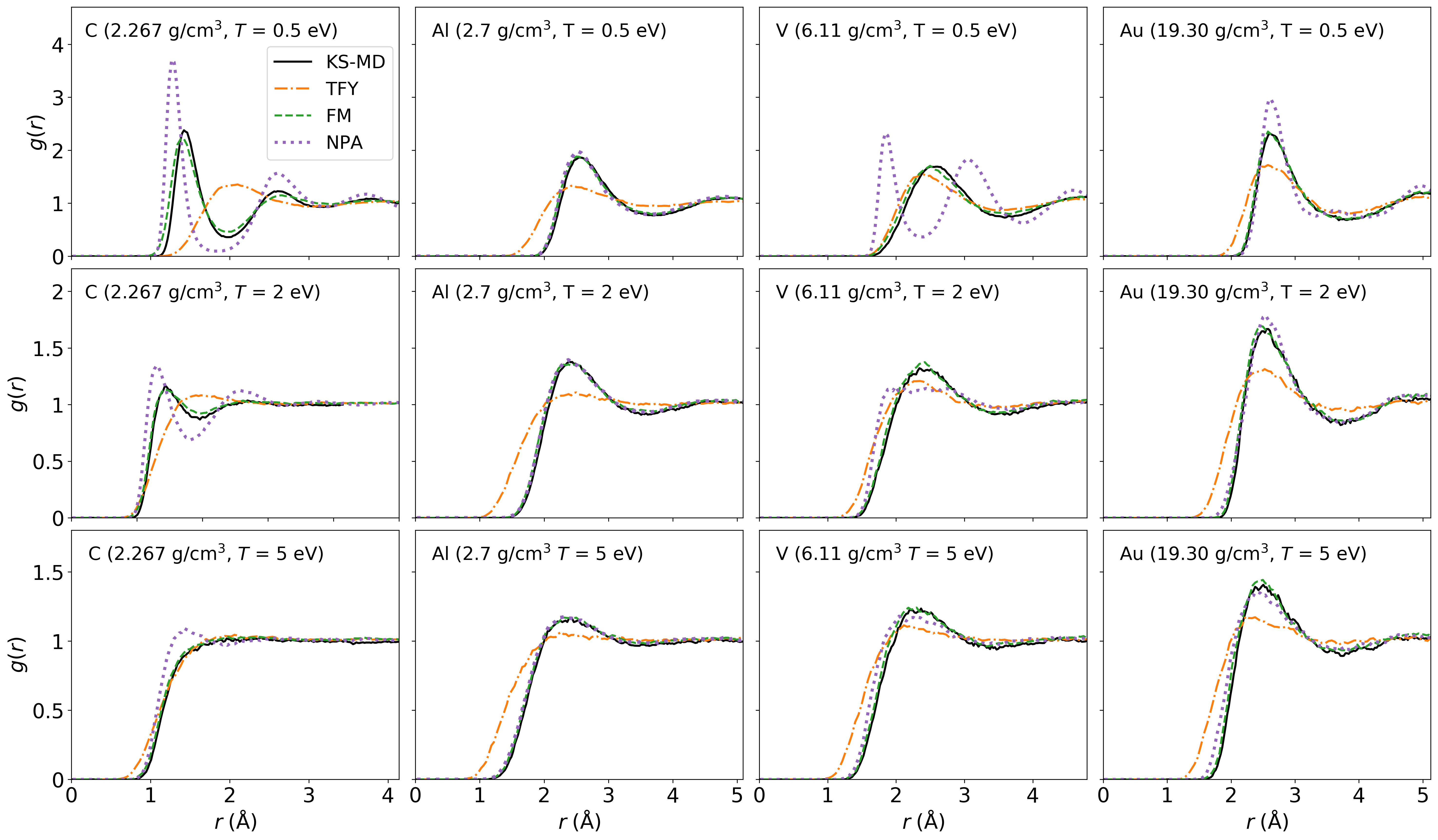}
    \caption{The radial distribution functions for C, Al, V, and Au are shown. The top row corresponds to $T = 0.5$ eV, the middle row $T = 2$ ev, and the bottom row $T = 5$ eV. The force-matched RPP always reproduces the $g(r)$ obtained from KS-MD.}
    \label{fig:gr_compare}
\end{figure*}

The radial distribution function measures the probability of finding another ion at a radial distance $r$ around a central ion. It has been shown that in general, there always exists a RPP that can reproduce $g(r)$ from a $N$-body simulation \cite{chihara1991unified} and the force-matching procedure provides a avenue for obtaining this RPP. Fig. \ref{fig:gr_compare} compares $g(r)$ computed from MD simulation for all RPP models for C, Al, V, and Au. Each row corresponds to a different temperature and clear trends can be observed, such as the improvement in agreement between models as the temperature increases. We note that the force-matched RPP always obtains the correct $g(r)$ and the NPA model generally predicts the location of the first peak but sometimes over-predicts the magnitude or misses the location of the first peak altogether as observed in the case of V at 6.11 g/cm$^3$ for $T = 0.5$ eV. The TFY model always underestimates the magnitude of the first peak height and the location is usually shifted.

\begin{figure*}
    \centering
    \includegraphics[width=17.2 cm]{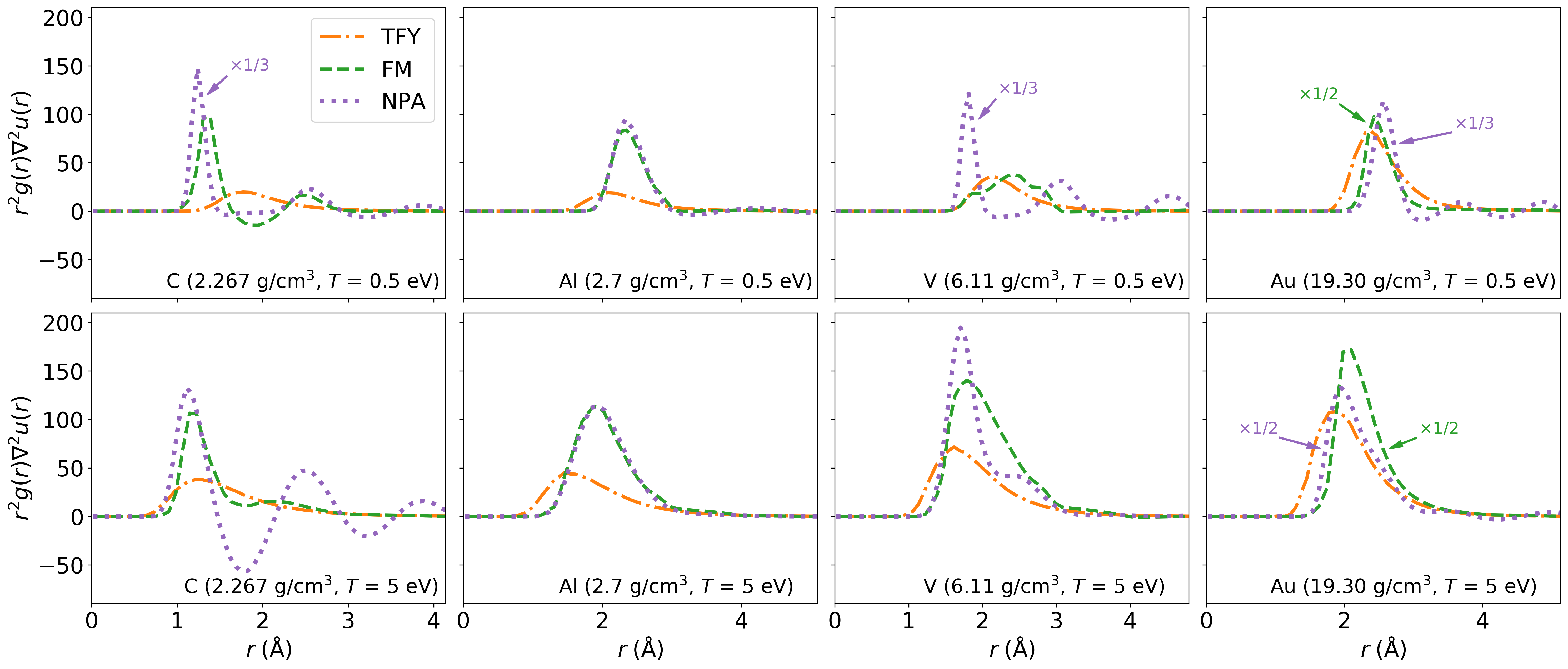}
    \caption{The integrand of the Einstein frequency (\ref{eq:omegasq}). All integrands are consistent with values reported in Table \ref{tab:transport} as the self-diffusion coefficient decrease as the integral of the Einstein frequency increases. This allows for a ``by eye" comparison of the self-diffusion coefficient from different RPP models.}
    \label{fig:EF_compare}
\end{figure*}

To understand how underestimating the $g(r)$ peak height impacts the self-diffusion coefficient, we use the Einstein frequency which is obtained through a short time expansion of the normalized velocity autocorrelation function
\begin{equation}
    Z(t) = 1 - \Omega_0^2 \frac{t^2}{2!} + \Omega_0^4 \frac{t^4}{4!} + \cdots, \label{eq:STE}
\end{equation}
where $\Omega_0$ is the Einstein frequency
\begin{equation}
    \Omega_0^2 = \frac{4\pi\rho_i}{3m_i}\int_0^\infty\! dr\: r^2 g(r) \nabla^2u(r),\label{eq:omegasq}
\end{equation}
where $m_i$ is the ion mass in grams. The Einstein frequency gives insight on the relationship between $u(r)$ and $g(r)$, highlighting how different regions are weighted more or less depending on the curvature of $u(r)$. In Fig. \ref{fig:EF_compare}, the integrand of (\ref{eq:omegasq}) is shown. For the TFY model, the integrand is always smaller than those predicted by force-matched and NPA RPPs. The area under each curve in Fig. \ref{fig:EF_compare} can be directly connected to the self-diffusion coefficient through the Green-Kubo relation (in 3 dimensions)
\begin{equation}
    D = \frac{T}{3m}\int_0^\infty\! dt\: Z(t),\label{eq:GK_D}
\end{equation}
by substituting, (\ref{eq:STE}) into (\ref{eq:GK_D}). Doing so shows that the TFY model will always predict a larger self-diffusion coefficient than the force-matched or NPA model as the area under these curves is larger. This is confirmed later when the self-diffusion coefficients are explicitly calculated as discussed in Sec. \ref{sec:SD}.

\subsection{\label{sec:SD}Self-Diffusion}

\begin{figure*}
    \centering
    \includegraphics[width=17.2 cm]{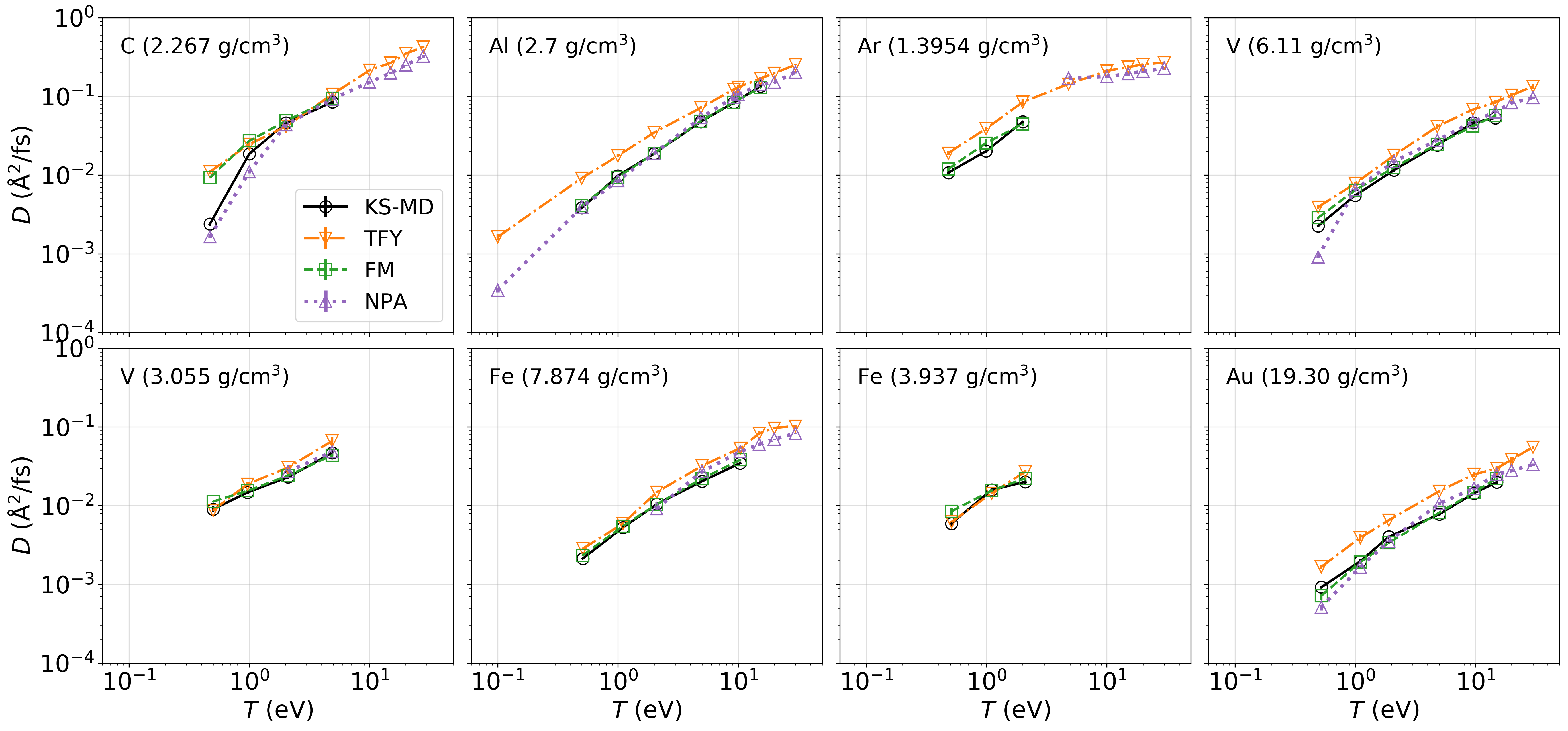}
    \caption{Self-diffusion coefficients for different elements and densities versus temperature. The numerical values are reported in Table \ref{tab:transport}. For all cases all models predict values that have roughly the same order of magnitude. The only case where the force-matched RPP fails to reproduce the KS-MD self-diffusion coefficient is for C at 2.267 g/cm$^3$ and $T = 0.5$ eV. General trends in the data include the over prediction in the self-diffusion coefficient from the TFY RPP, consistent with the Einstein frequency in Fig. \ref{fig:EF_compare}. This over prediction is reduced greatly when using a NPA RPP and is generally in good agreement with the force-matched RPP and KS-MD data.} 
    \label{fig:SD_compare}
\end{figure*}

One approach to compute the self-diffusion coefficient is via the the slope of the mean-squared displacement from the Einstein relation
\begin{equation}
    D = \lim_{t \to \infty}\frac{\langle|\mathbf{r}(t) - \mathbf{r}(0)|^2\rangle}{6t}, \label{eq:msd}
\end{equation}
where $\langle \cdot \rangle$ is an ensemble average over particles {\it{and}} time. Relative to (\ref{eq:GK_D}) we have found that (\ref{eq:msd}) converges faster for small numbers of particles and short simulation lengths, and is less sensitive to finite-size effects. Therefore, all self-diffusion coefficients reported in this work have been calculated from a linear fit to the mean-squared displacement, $\langle|\mathbf{r}(t) - \mathbf{r}(0)|^2\rangle$. 

Due to finite-size effects, two problems arise when computing the slope and uncertainty of the linear fit. First, we must ensure that the linear fit is carried out in the late-time linear regime of the mean-squared displacement. Second, we dismiss statistically unconverged late-time behavior of the mean-squared displacement where the ensemble average contains sparse amounts of data. To remedy both of these concerns, we uniformly randomly sub-sample the mean-squared displacement 100 times with 10 points along each sub-sample. Next, a linear fit is determined for each sub-sample and the standard deviation of the sub-sample slopes is computed. Once the standard deviation is known, a cutoff time is calculated by determining the point in time that the standard deviation of the sub-sample fits is less than half of the standard deviation computed from {\it{sub-sample}} fits to the {\it{entire}} mean-squared displacement. The simulation data for the mean-squared displacement after the cutoff time is discarded and the fitting procedure described above is repeated. The average and standard deviation of the new sub-sample fits yields self-diffusion coefficient and the uncertainty in the fit respectively and are reported in Table \ref{tab:transport}. 

\begin{figure}
\includegraphics[width=8.6 cm]{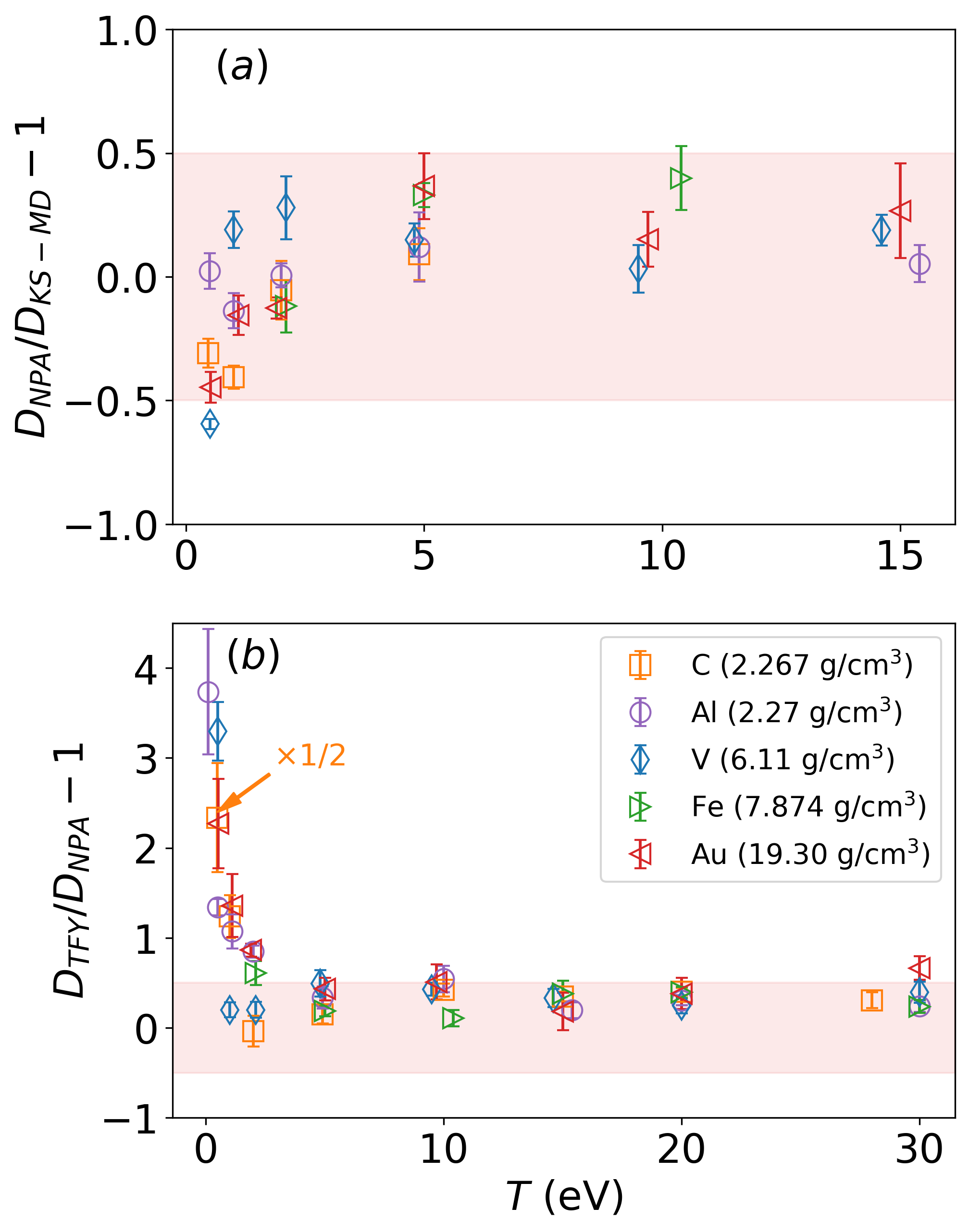} 
\caption{Relative self-diffusion coefficients. The shaded region brackets the range of $-0.5$ and $0.5$. $(a)$ For all cases except vanadium at $T = 0.5$ eV, the points fall within the bounds of the bracketed region. The accuracy of the NPA model for low temperatures suggests where KS-MD is most advantageous. In $(b)$ the points are within 50\% of the NPA value from $T = 5$ eV and above for most cases. The orange point marked with an arrow has been reduced by a factor of $1/2$ to improve clarity of the banded region.}
\label{fig:SD_ratio}
\end{figure}

By computing the relative self-diffusion coefficients reported in Table \ref{tab:transport} to determine a rule where NPA is accurate relative to KS-MD and similarly where TFY is accurate relative to NPA. Doing so, we are informed of when certain models may be accurate and when others are not. For example, the top panel in Fig. \ref{fig:SD_ratio} suggests that NPA models may be accurate from $T = 1$ eV and above if the target error tolerance is 50\% of the self-diffusion coefficient computed from KS-MD. Similarly in the bottom figure, the TFY model is generally accurate to within 50\% of the NPA model from $T = 5$ eV and beyond.

\begin{table*}
    \caption{The self-diffusion coefficient for all systems. For each RPP model, the number of particles, time step, and simulation length were kept identical for each element, density, and temperature. Finite-size corrections are carried out in Sec. \ref{sec:FSE}.}
    \begin{ruledtabular}\label{tab:transport}
        \begin{tabular}{l l l l l l l}
        Element & $\rho_i$ (g/cm$^3$) & $T$ (eV) & $D_{KS-MD} \; (10^{-3} \text{\AA}^2/\text{fs})$ &  $D_{FM} \; (10^{-3}\text{\AA}^2/\text{fs})$ & $D_{TFY} \; (10^{-3} \text{\AA}^2/\text{fs})$  & $D_{NPA} \; (10^{-3} \text{\AA}^2/\text{fs})$\\
        \hline\\
        Li & 0.513 & 0.054 & 1.4 $\pm$ 0.13 & 1.27 $\pm$ 0.054 & 5.6 $\pm$ 0.39 & 1.26 $\pm$ 0.077\\\\
        C & 2.267 & 0.47 & 2.4 $\pm$ 0.12 &  9.3 $\pm$ 0.20 & 11.0 $\pm$ 0.60  &  1.69 $\pm$ 0.060\\
        &  & 1.0 & 18.6 $\pm$ 0.7 &  27.2 $\pm$ 0.77 &   25 $\pm$ 1.64  &  11.0 $\pm$ 0.45  \\
        &  & 2.0 & 46 $\pm$ 1.68 & 49 $\pm$ 1.0 & 42 $\pm$ 3.70  & 43 $\pm$ 3.86  \\
        &  & 4.9 & 85 $\pm$ 5 & 94 $\pm$ 5.82 & 106 $\pm$ 5.37 & 92 $\pm$ 3.45 \\
        &  & 10.0   & -- & -- & 215 $\pm$ 4.49 &  151 $\pm$ 4.63\\
        &  & 15   & -- & -- & 266 $\pm$ 3.46 & 198 $\pm$ 5.10\\
        &  & 20   & -- & -- & 349 $\pm$ 7.60 & 249 $\pm$ 2.35\\
        &  & 28   & -- & -- &  423 $\pm$ 13.28 & 324 $\pm$ 12.17 \\\\
        Al & 2.7 & 0.1 & -- & -- & 1.6 $\pm$ 0.14 &  0.35 $\pm$ 0.0217 \\
        &  & 0.50 & 3.8 $\pm$ 0.16 & 4.1 $\pm$ 0.13 & 9.17 $\pm$ 0.099  & 3.9 $\pm$ 0.11 \\
        &  & 1.1 &  9.8 $\pm$ 0.30 &  9.4 $\pm$ 0.11 & 17.5 $\pm$ 0.70  & 8.5 $\pm$ 0.44\\
        &  & 2.0 & 18.7 $\pm$ 0.50 & 18.8 $\pm$ 0.68 & 34.8 $\pm$ 0.52  &  18.8 $\pm$ 0.40\\
        &  & 4.9 & 48 $\pm$ 3.56 & 49 $\pm$  3.17 & 72 $\pm$ 3.03  & 54 $\pm$ 2.7\\
        &  & 9.2 & 83 $\pm$ 1.63 & 84 $\pm$ 5.67 & 122 $\pm$  5.83 & 94 $\pm$ 8.80\\
        &  & 10.0 & -- & -- &  131 $\pm$ 6.77  & 105 $\pm$ 13.53\\
        &  & 15.4 &  134 $\pm$ \; 3.68 &  129 $\pm$ 3.37 &  169 $\pm$ 5.26 &  142 $\pm$ 6.17 \\
        &  & 20.0 & -- & -- & 197 $\pm$ 5.21 & 151 $\pm$ 2.55 \\
        &  & 30.0 & -- & -- & 252 $\pm$ 4.74 & 203 $\pm$ 6.44 \\\\
        Ar & 1.395 & 0.48 & 10.7 $\pm$ 0.43 & 12 $\pm$ 1.03 & 19 $\pm$ 1.10  & --\\
        &  & 1.0 &  20.1 $\pm$  0.89 & 26 $\pm$ 3.0 &  39 $\pm$ 2.22 & -- \\
        &  & 2.0 & 48 $\pm$ 1.75 & 45 $\pm$ 2.84 & 85 $\pm$ 8.75 & -- \\
        &  & 5  & -- & -- & 143 $\pm$ 6.55 &  171 $\pm$ 4.09\\
        &  & 10.0 & -- & -- & 210 $\pm$ 14.53 & 179 $\pm$ 6.21 \\
        &  & 15.0 & -- & -- & 235 $\pm$ 13.34 & 193 $\pm$  11.95 \\
        &  & 20.0 & -- & -- & 255 $\pm$ 6.73 & 209 $\pm$ 8.91 \\
        &  & 30.0 & -- & -- & 268 $\pm$ 2.73 & 228 $\pm$ 8.26 \\\\
        V & 6.11 & 0.49 & 2.25 $\pm$ 0.050  & 2.86 $\pm$ 0.079 & 3.9 $\pm$ 0.18 & 0.91 $\pm$ 0.027\\
        &  & 1.0 &  5.5 $\pm$ 0.21 & 6.5 $\pm$ 0.16 & 7.9 $\pm$ 0.36 &  6.6 $\pm$ 0.15\\
        &  & 2.1 &  11.6 $\pm$ 0.78 & 12.5 $\pm$ 0.68 & 17.8 $\pm$ 0.74 & 14.8 $\pm$ 0.50\\
        &  & 4.8 & 24.2 $\pm$ 0.63 & 24.7 $\pm$ 0.88 & 41 $\pm$ 2.76 & 27.7 $\pm$ 0.90\\
        &  & 9.5 & 46 $\pm$  3.41 & 42 $\pm$  2.65 & 68 $\pm$ 2.10 & 47.6 $\pm$ 0.93 \\
        &  & 14.6 &  53 $\pm$ 1.81 & 57 $\pm$ 3.25 & 84 $\pm$ 4.83 & 63 $\pm$ 1.19\\
        &  & 20.0 & -- & -- &  103 $\pm$ 6.10 &  82.7 $\pm$ 0.78 \\
        &  & 30.0 & -- & -- & 134 $\pm$ 8.57 & 96 $\pm$ 1.86\\
        &  3.055 & 0.5 & 9.0 $\pm$ 0.81 & 11.3 $\pm$ 0.29 &  8.7 $\pm$ 0.23 & --\\
        &  & 0.97 &  14.7 $\pm$ 0.47 & 15.4 $\pm$ 0.43 & 19 $\pm$ 1.39 & --\\
        &  & 2.0 & 23 $\pm$ 1.13 & 24 $\pm$ 1.84 & 31 $\pm$ 1.26 & 27 $\pm$ 1.84\\
        &  & 4.9 &  47 $\pm$ 4.38 & 44 $\pm$ 2.52 & 66 $\pm$ 7.30 & 48 $\pm$ 2.02\\\\
        Fe & 7.874 & 0.51 & 2.13 $\pm$ 0.047  & 2.34 $\pm$ 0.042 & 2.84 $\pm$ 0.030 & -- \\
        &  & 1.1 & 5.27 $\pm$ 0.098 & 5.5 $\pm$ 0.16 &  5.9$\pm$ 0.39 & -- \\
        &  & 2.1 & 10.4 $\pm$ 0.72 & 10.4 $\pm$ 0.73  & 14.8 $\pm$ 0.46 & 9.2 $\pm$ 0.47\\
        &  & 5.0 & 20.4 $\pm$ 0.61 & 22.0 $\pm$ 0.97 & 32 $\pm$ 1.41 & 27.1 $\pm$ 0.19\\
        &  & 10.4 & 35 $\pm$ 1.14 & 38 $\pm$ 1.40 & 54 $\pm$ 1.25 & 49 $\pm$ 2.90\\
        &  & 15.0 & -- & -- & 83 $\pm$  5.18 &  60 $\pm$ 2.60\\
        &  & 20.0 & -- & -- & 97 $\pm$ 2.93 & 70 $\pm$ 1.04 \\
        &  & 30.0 & -- & -- & 103 $\pm$ 4.69 & 83.0 $\pm$ 0.94 \\
        &  3.937 & 0.51 & 6.0 $\pm$ 0.39 & 8.5 $\pm$ 0.94 & 6.2 $\pm$ 0.21 & -- \\
        &  & 1.1 & 15.8 $\pm$ 0.70 &  15.6 $\pm$ 0.67 &  14.4 $\pm$ 0.33 & --\\
        &  & 2.1 & 20 $\pm$ 1.18 & 22 $\pm$ 2.07 & 27 $\pm$ 1.28 & --\\\\
        Au & 19.30 & 0.52 &  0.92 $\pm$ 0.028 & 0.71 $\pm$ 0.084 & 1.67 $\pm$ 0.12 &  0.51 $\pm$ 0.042\\
        &  & 1.1 & 2.0 $\pm$ 0.11 & 1.92 $\pm$ 0.088 & 3.9 $\pm$ 0.42 & 1.66 $\pm$ 0.069\\
        &  & 1.9 & 4.0 $\pm$ 0.14 & 3.4 $\pm$ 0.15 & 6.6 $\pm$ 0.16 & 3.52 $\pm$ 0.05\\
        &  & 5.0 & 7.8 $\pm$ 0.40 & 8.2 $\pm$ 0.21 & 15.3 $\pm$ 0.63 &  10.7 $\pm$ 0.50\\
        &  & 9.7 & 14.4 $\pm$ 0.64 & 15 $\pm$ 1.19 & 25 $\pm$ 1.94 &  16.6 $\pm$ 0.86 \\
        &  & 15.0 &  19.82 $\pm$ 0.80 & 22 $\pm$ 2.56 & 30 $\pm$ 1.95 & 25 $\pm$ 2.79\\
        &  & 20.0 &  -- & -- & 39 $\pm$ 3.49 &  28.0 $\pm$ 0.97\\
        &  & 30.0 &  -- & -- &  56 $\pm$ 2.23 & 33 $\pm$ 1.26\\
        \end{tabular}
    \end{ruledtabular}
\end{table*}

Two important observations can be made from the trends in Fig. \ref{fig:SD_ratio}. The top panel illustrates temperatures at which an $N$-body potential is needed and when NPA is adequate. The bottom panel shows a comparison with TFY, which has the simplest $u_{ei}(k)$ and $\chi(k)$, and we see temperatures at which TFY becomes comparable to NPA, suggesting when we can exploit simpler approximations for those inputs.

\subsection{Power Spectrum}

\begin{figure*}
    \centering
    \includegraphics[width=17.2 cm]{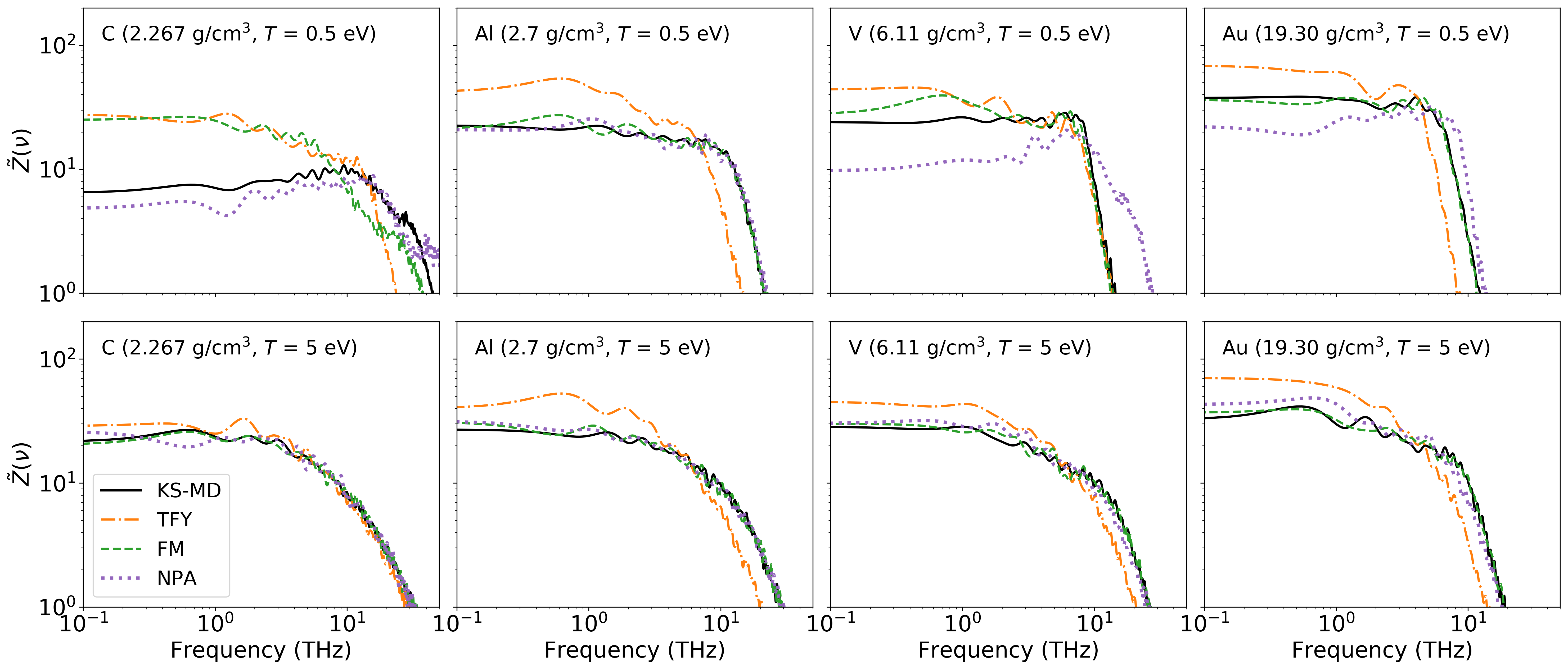}
    \caption{The normalized power spectrum for C, Al, V, and Au. For C at $T = 0.5$ eV, the single particle dynamics are poorly described by the TFY and force-matched models but more accurately described with the NPA model. As the temperature increases from $T = 0.5$ to $5$ eV, all models more accurately reproduce small and high frequency dynamics with the most notable improvement for C.}
    \label{fig:Zw_compare}
\end{figure*}

The self-diffusion coefficient is useful for comparing and quantifying the accuracy of RPP models and transport theories, but in order to assess how accurately the particle dynamics are reproduced, we look at the power spectrum of the velocity autocorrelation function $Z(t)$ 
\begin{equation}
    \tilde{Z}(\nu) = \int_{0}^{\infty} dt \cos{\left( 2\pi\nu t\right)Z(t)}.
\end{equation}
In Fig. \ref{fig:Zw_compare}, we compare $\tilde{Z}(\nu)$ calculated using TFY, force-matched, and NPA RPPs against results obtained from KS-MD. We find that with the exception of low temperature C and V, force-matched RPPs agree with the KS-MD results across the entire frequency range.  This, combined with the low relative force errors and accurate reproduction of static properties discussed previously, indicates that the the force-matched RPPs accurately approximate the Kohn-Sham potential energy surface. For higher temperatures, the NPA RPP is very similar to the force-matched RPP for low and high frequencies for all elements. For $T = 0.5$ eV, the dynamics predicted from the NPA model are noticeably less similar to those from KS-MD where NPA underestimates the prevalence of low-frequency modes in Au and both low and high-frequency modes in V. Interestingly, the NPA RPP captures the single-particle dynamics of low temperature C very well, but Figs. \ref{fig:ur_compare} and \ref{fig:gr_compare} indicate that this is agreement comes at the expense of sacrificing the accuracy of static properties. Lastly, the TFY RPP exhibits roughly the same trends across all elements and temperatures-- overestimation of the low frequency modes and underestimation of the high-frequency modes except for the case of C at 2.267 g/cm$^3$ and $T = 5$ eV where excellent agreement with KS-MD is observed. 

\subsection{\label{sec:FSE}Finite-Size Corrections}
\begin{figure}
    \centering
    \includegraphics[width=8.6 cm]{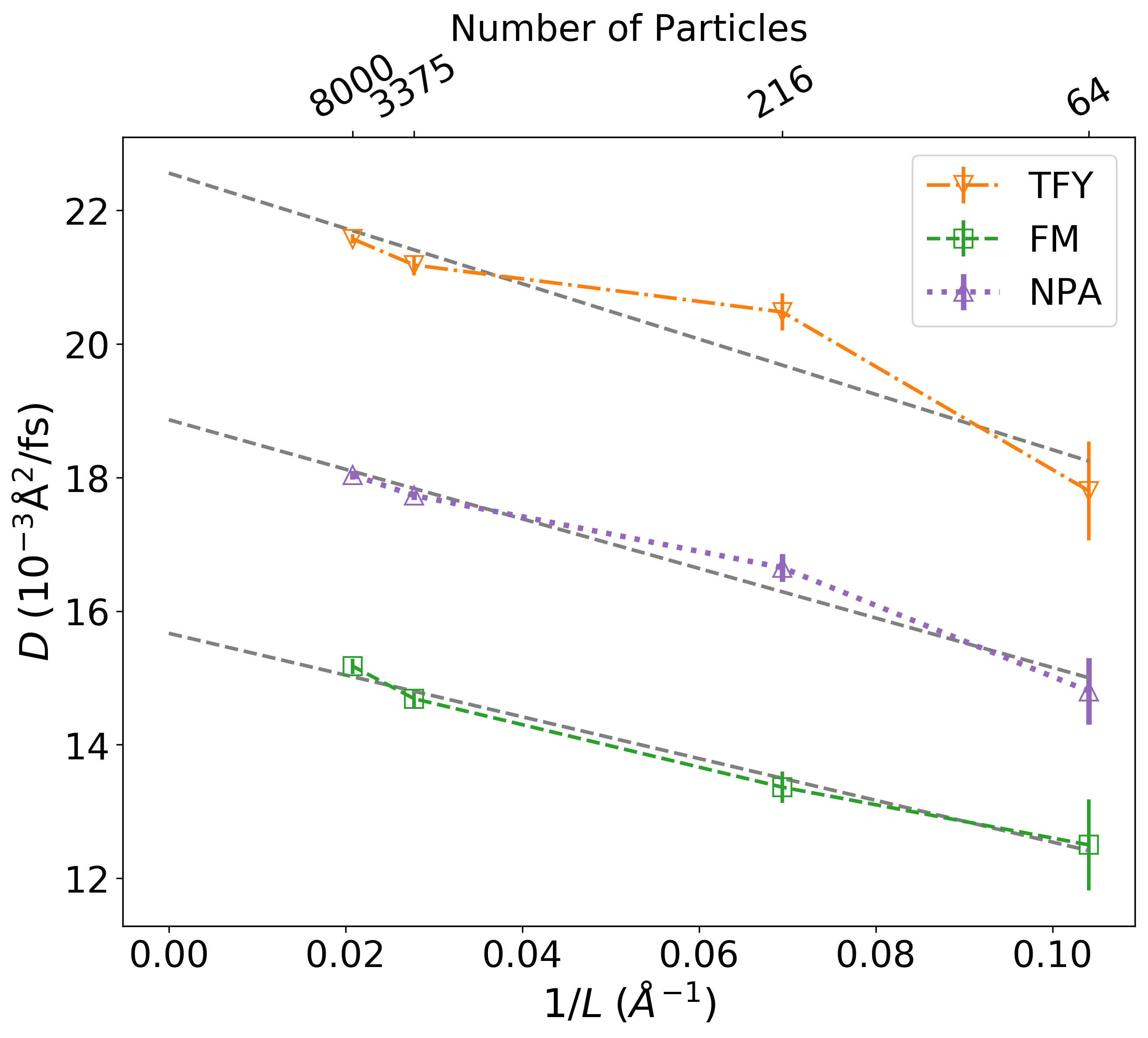}
    \caption{Finite-size effect study for V at 6.11 g/cm$^3$ and $T = 2$ eV. Identical MD simulations were carried out with increasing particle number. Extrapolating with a linear fit (grey dashed line) to $1/L = 0$ approximates the thermodynamic limit, correcting the values in Table \ref{tab:transport}.}
    \label{fig:fse_fit}
\end{figure}

Generally, thousands or even millions of atoms are needed to approximate the thermodynamic limit \cite{germann2008trillion, stanton2018multiscale}. While the KS-MD framework provides an accurate description of the electronic structure and the $N$-body potential is determined on-the-fly, corrections for finite-size effects must be considered. When the shear viscosity $\eta$ of the system is known, finite-size corrections can be determined from \cite{finite-size}
\begin{equation}
    D_{\infty} = D_N + \frac{\xi T}{6\pi \eta L},
\end{equation}
 where $D_\infty$ is the self-diffusion in the thermodynamic limit, $D_N$ is the self-diffusion coefficient computed from a system of finite number of particles $N$, and $\xi = 2.837297$ for cubic simulation boxes with periodic boundary conditions. When $\eta$ is unknown, multiple simulations of increasing particle number are carried out and a linear fit is used to determine $D_\infty$. Results from this procedure are shown in Fig. \ref{fig:fse_fit} where $D_\infty$ is determined via linear extrapolation to $1/L = 0$. 
 
 By finding the percent difference in $D_\infty$ and $D_N$, we approximate the errors from finite-size effects in the KS-MD self-diffusion coefficient at these conditions. The approximate error in KS-MD for the case shown in Fig. \ref{fig:fse_fit}, is $\sim 20$\%. While the error will vary with $\{Z, n, T\}$, the impact of finite-size effects is significant. From this study, the most promising approach is to fully converge the NPA MD results, using force-matched RPPs when necessary (for low temperatures $T \lesssim 1$ eV).

 Finite-size corrections allow for a direct comparison to analytic transport theories, namely the Stanton-Murillo model \cite{SMT_paper}. The Stanton-Murillo model, provides a closed form solution for ionic self-diffusion by using an effective interaction potential in a Boltzmann kinetic theory framework.
 The major benefit of this model is that the computation of ionic transport is nearly instantaneous. However, its applicability in the cold dense matter and warm dense matter regimes is unknown. 

The results in Table \ref{tab:fse_transport} show that the effective interaction approach of the Stanton-Murillo model captures much of the many-body physics included in the TFY RPP results. The main weakness of the model, and also TFY, is therefore the functional form of the interaction they employ, as the differences with the force-matched and NPA columns reveal. Because self-diffusion is a relatively simple transport coefficient \cite{SMT_paper}, more work is needed to quantify these trends for other transport properties.

\begin{table*}
    \caption{Self-diffusion coefficients in the thermodynamic limit. Both elements are at solid density (2.267 g/cm$^3$ for C, and 6.11 g/cm$^3$ for V).}
    \begin{ruledtabular}\label{tab:fse_transport}
        \begin{tabular}{l l l l l l l }
        Element & $T$ (eV) & $D_{FM} \; (10^{-3} {\text{\AA}}^2/{\text{fs}})$ & $D_{NPA} \; (10^{-3} {\text{\AA}}^2/{\text{fs}})$ &  $D_{TFY} \; (10^{-3} {\text{\AA}}^2/{\text{fs}})$  & $D_{SM} \; (10^{-3} {\text{\AA}}^2/{\text{fs}})$ & $D_{CSM} \; (10^{-3} {\text{\AA}}^2/{\text{fs}})$\\
        \hline\hline
        C  & 0.47 &  10.55 & 2.14  &  12.66 & 13.08&  2.14 \\
        &   1.0  & 32.44   & 14.21   & 25.57  &  26.11  & 13.87 \\
        &   2.0  & 56.70   &  43.12  & 51.14  &  50.53 & 39.23\\
        &   4.9  &  99.51  &  109.55  & 117.88  & 118.34 &  106.76\\
        &   10.0 & -- & 169.91 &  210.76  &  217.34 &  206.71 \\
        &   15.0   & -- & 219.99 &  296.21 & 293.54 & 284.10\\
        &   20.0   & -- &  256.33 &   342.15 & 356.41 &  348.04 \\
        &   28.0   & -- & 327.32&  470.10 & 439.44 & 432.07 \\\hline
        V  & 0.49 & 4.14  & 1.01 &  5.42 & 6.76 & 4.39\\
        &  1.0   & 8.54  & 8.53 & 11.53 & 12.26 & 7.96\\
        &  2.1   & 15.67 & 18.87 & 22.56 & 23.14 & 15.03\\
        &  4.8   & 28.72 & 31.34 & 42.42 & 46.25  & 30.18\\
        &  9.5   & 49.49 & 54.90 & 73.76 & 77.10  & 51.16\\
        &  14.6  & 66.98 &  74.44 & 99.82 & 99.78  & 67.97\\
        &  20.0    & --    & 87.90 &  118.24 & 117.89  & 83.15\\
        &  30.0    & --    & 105.60 & 143.63 & 141.66 & 106.94\\
        &  50.0    & --    &  131.84 &  175.35 & 171.07 & 143.02\\
        &  75.0    & --    &  178.34 & 202.93 & 194.31 & 172.91\\
        &  100.0   & --    & 209.50 & 207.20 & 211.86 & 194.36
        \end{tabular}
    \end{ruledtabular}
\end{table*}

\begin{figure}
    \centering
    \includegraphics[width=8.6 cm]{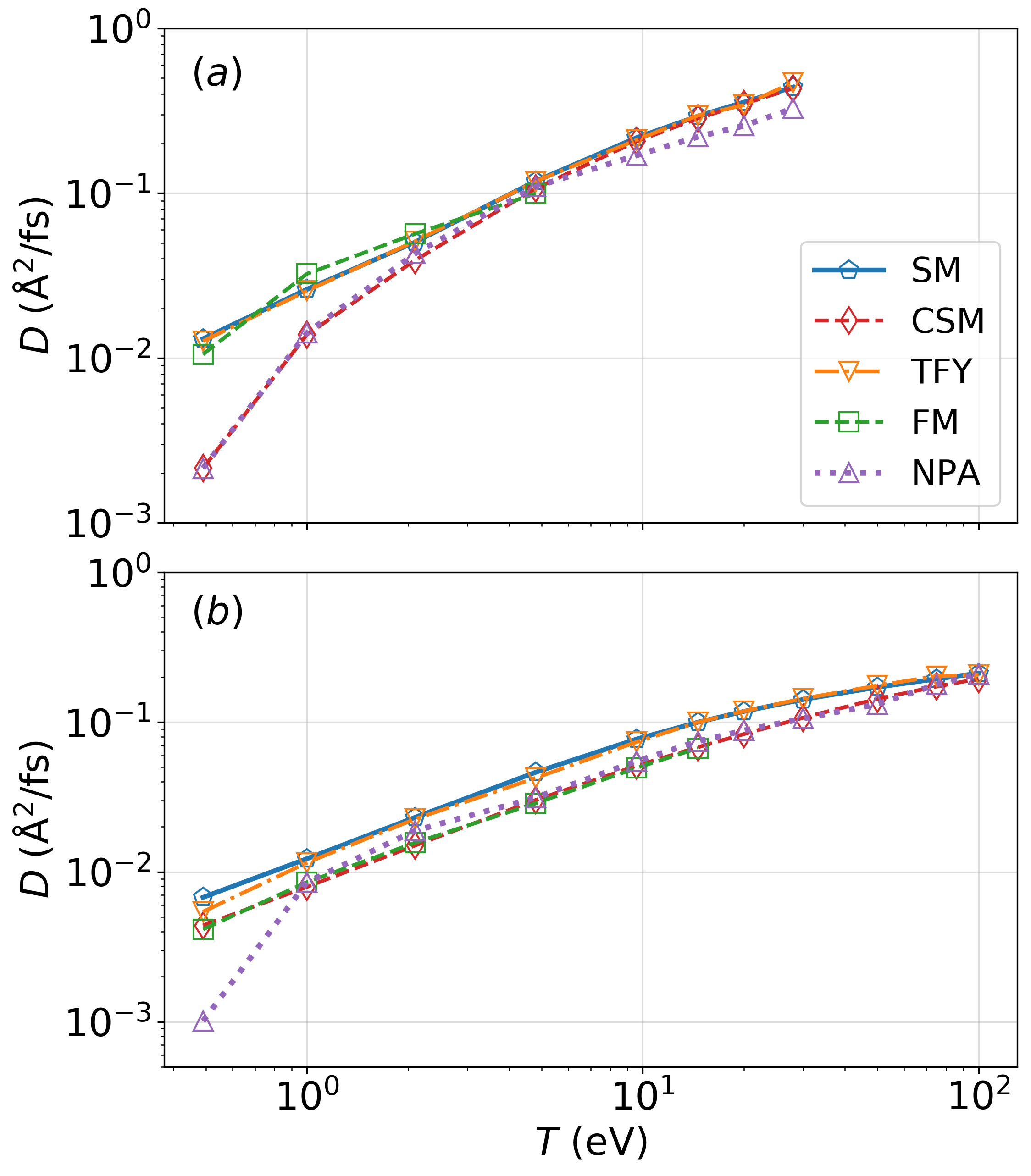}
    \caption{Self-diffusion coefficient versus temperature in the thermodynamic limit. The points displayed here are taken from Table \ref{tab:fse_transport}. $(a)$ Self-diffusion coefficient for C at 2.267 g/cm$^3$.$(b)$ Self-diffusion coefficient for V at 6.11 g/cm$^3$. The Stanton-Murillo model (denoted SM) fails for low temperature C. For V, the Stanton-Murillo model shows excellent agreement with the force-matched RPP even at low temperatures. The validity of the Stanton-Murillo model is extended to low temperatures with an effective interaction correction (denoted CSM).}
    \label{fig:fse_compare}
\end{figure}

With the converged self-diffusion data, we generate an effective interaction correction to original Stanton-Murillo model. The effective interaction corrected Stanton-Murillo model is
\begin{equation}
    D_{CSM} = \alpha(Z, T) D_{SM},
\end{equation}
where $\alpha(Z,T)$ is determined by fitting the ratio of the self-diffusion coefficient from the best performing RPP model and the self-diffusion coefficient computed from the Stanton-Murillo model $D_{SM}$ to the functional form
\begin{equation}
    \alpha(Z,T) = \frac{a \text{erf}(bT)}{bT} + 1,\label{eq:smt_corr}
\end{equation}
which asymptotes to $D_{SM}$ as $T$ increases. Here the ``best performing RPP model" refers to the RPP model that most accurately reproduced the self-diffusion coefficient computed from 64 particle KS-MD simulations. The parameters $a$ and $b$ are reported in Table \ref{tab:correc_coeff} for C at 2.267 g/cm$^3$ and V at 6.11 g/cm$^3$ and their values vary considerably between both cases emphasizing the need for a comprehensive finite size effect study  to produce correction factors for additional elements and conditions. This correction factor allows for the use of the Stanton-Murillo model in regions of previously unknown accuracy. The finite-size corrections along with the corrected Stanton-Murillo model results are shown Fig. \ref{fig:fse_compare} with the numerical values given in Table \ref{tab:fse_transport}. Note that for low temperature C at 2.267 g/cm$^3$, the best performing RPP model was NPA (as reported in Fig. \ref{fig:SD_compare} and Table \ref{tab:transport}) explaining why the corrected Stanton-Murillo model tends towards the NPA RPP at low temperatures. For V at 6.11 g/cm$^3$ the best performing RPP model was the force-matched RPP again explaining the low temperature trend.

\begin{table}
    \caption{Coefficients $a$, and $b$ for the effective interaction correction (\ref{eq:smt_corr}). Note that the values of $a$ and $b$ vary considerably for each element.}
    \begin{ruledtabular}\label{tab:correc_coeff}
        \begin{tabular}{l l l}
        Element & $a$ & $b$ \\
        \hline
        C (2.267 g/cm$^3$) & 2.198 &  -1.032  \\
        V (6.11 g/cm$^3$) & 0.03767 & -0.3112
        \end{tabular}
    \end{ruledtabular}
\end{table}

\begin{figure}
\includegraphics[width=8.6 cm]{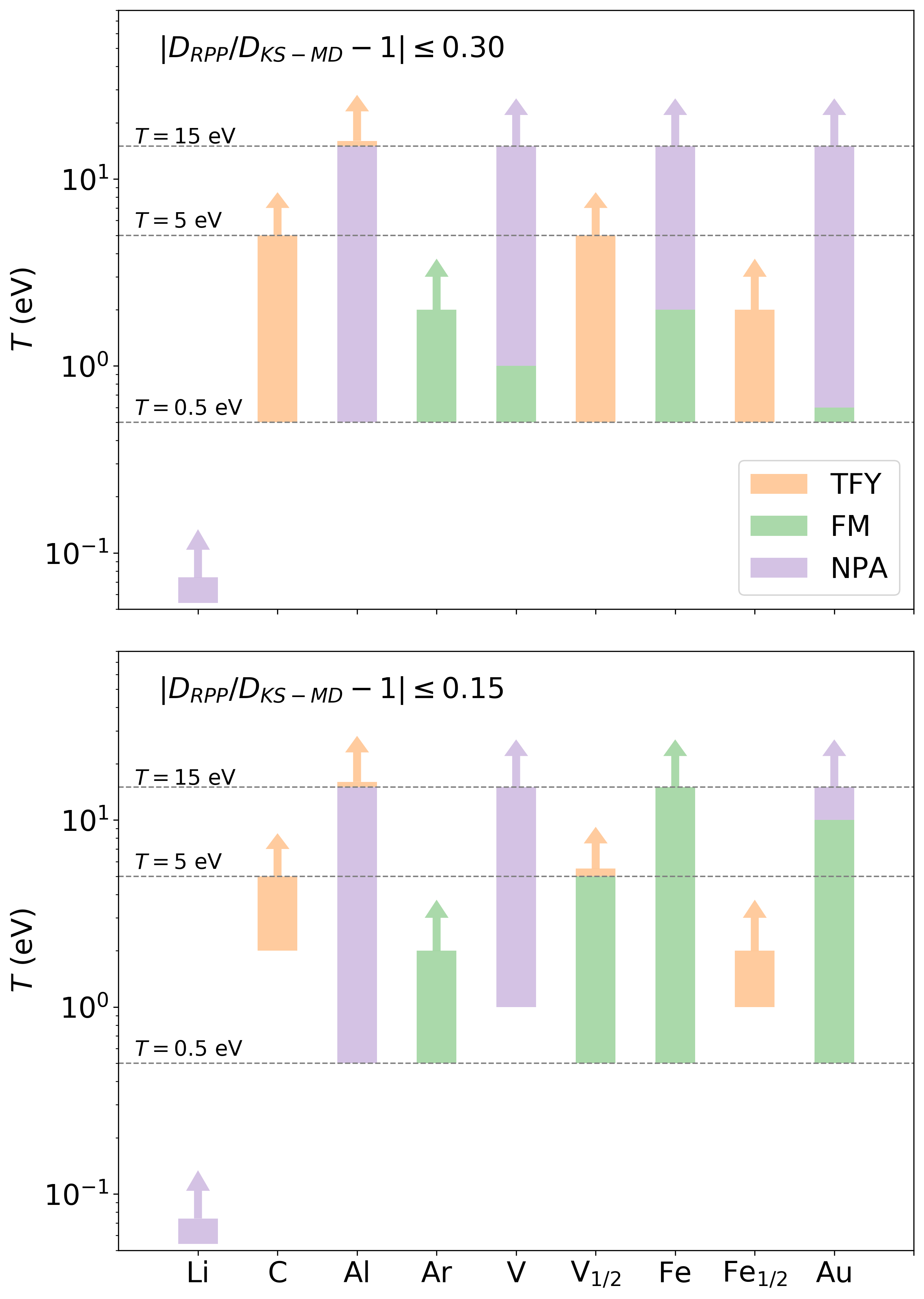} 
\caption{Suggested use cases for each RPP model based on the relative self-diffusion coefficient error (between RPP-MD and KS-MD) and cheapest computation cost. The top and bottom panels correspond to a 30\% and 15\% relative error respectively. The elements denoted with a subscript  of ``1/2" corresponds to half solid density (V at 3.055 g/cm$^3$ and Fe at 3.937 g/cm$^3$). The colored bars indicate the computationally cheapest RPP that generates a self-diffusion coefficient to within the specified error tolerance available for that system based on Table \ref{tab:transport}. The empty space under each bar indicates regions where no KS-MD data was collected so no assessment on a RPPs accuracy can be made.}
\label{fig:model_efficacy}
\end{figure}

In an attempt to summarize our work in a single figure, Fig. \ref{fig:model_efficacy} shows our suggested use cases for all RPPs studied here for two relative self-diffusion accuracies computed from  Table \ref{tab:transport}. When points (the average value or its uncertainty) for a given model are within the appropriate tolerance (30\% for the top panel and 15\% for the bottom panel), we consider the model as being accurate for that temperature and element and is denoted with a colored bar or arrow. We rank the computational expense from lowest to highest as: TFY, NPA, force-matching, and KS-MD. When a computationally cheaper model is accurate, it replaces the more computationally expensive model in Fig. \ref{fig:model_efficacy}. Based on trends observed in Figs. \ref{fig:f_err}, \ref{fig:SD_ratio}, and \ref{fig:fse_compare}, we assume that a models remain accurate for higher temperatures and is illustrated by the upward pointing colored arrows. Consider the case of Fe in the top panel of Fig. \ref{fig:model_efficacy}. The force-matched RPP is accurate to within 30\% of the KS-MD result from T = $0.5$ eV and up. The NPA model, which is computationally cheaper than the force-matched RPP, becomes accurate (within 30\% of KS-MD) at $T = 2$ eV and up. Hence the transition between the force-matched and NPA models. For Al, the NPA RPP is within 15\% of KS-MD at all temperatures. However, at $T = 15$ eV the TFY model becomes accurate therefore replacing the NPA RPP.

\section{\label{sec:conc}Conclusions and Outlook}

A systematic study of various RPPs for molecular dynamic simulations of dense plasmas was performed for a wide range of elements versus temperature for solid and half-solid density cases. Of the RPPs studied in this work, RPPs constructed from a NPA approach come closest to accurately reproducing the transport and structural properties predicted by KS-MD. The failures of NPA for metals near $T = 0.5$ eV are expected: V is a polyvalent metal and s-d hybridization occurs in Au, which are not treated well in our variant of the NPA model. Thus, it is unclear if inaccuracies in NPA reveal the need for $N$-body interactions or an improved NPA treatment. Moreover, finite-size corrections to KS-MD are seen to be significant; prior work on Silicon suggests that at least $108$ particles are needed to accurately treat elements like C at low temperatures \cite{remsing2017dependence}. Although this work does not fully resolve these issues, the trends seen for the lowest temperature for C, V and Au should be examined in detail in future work. 

As in previous work \cite{wunsch2009ion, White}, the TFY model tends to underestimate the peak (relative to KS-MD) of $g(r)$ at low temperatures, with a corresponding error in the self diffusion coefficient. Notionally the accuracy of the TFY model appears to follow the machine learning trend of $\langle Z \rangle/Z > 0.35$ \cite{murillo2020machine}, although it was not possible to use all models at high enough temperatures to be quantitative. In contrast, the NPA model with its improved Kohn-Sham treatment and use of a pseudopotential in (\ref{sopt}) eliminates most of these errors except for C and V at $T = 0.5$ eV, elements for which we would recommend NPA for $T > 2$ eV. Because we examined seven diverse elements over the warm dense matter regime, The accuracy of NPA (and for moderate temperature, even TFY) suggests that no additional ``short-range repulsion" \cite{wunsch2009ion, srr_nature, PhysRevLett.110.065001, Glenzer_2016, ruter2014ab, PhysRevLett.103.245004, doi:10.1016/j.mre.2017.09.001} is needed beyond (\ref{sopt}); as (\ref{sopt}) does not contain core-core repulsion, the structure of the interaction is more likely to be effective core-valence repulsion captured by $u_{ei}(k)$, as well as structure in $\chi(k)$ beyond $\chi_{TF}(k)$. In fact, our results suggest that core-core repulsion is quite small as ions do not approach each other on core length scales because of Coulomb repulsion; at higher temperatures, where ions could have closer encounters, the core is smaller; thus, the {\it first} term in (\ref{sopt}) is purely Coulombic.  

As expected, the force-matched potential reproduced the $g(r)$ computed from KS-MD for all cases. In only one case, again C at 2.267 g/cm$^3$ and $T = 0.5$ eV, the force-matched RPP overestimated the self diffusion coefficient; this suggests that the spherical pair interaction isn't applicable, and non-spherical corrections, which could include three-body contributions, are needed as suggested by the near-perfect agreement of the SNAP and KS-MD microfield of force magnitudes in Fig. \ref{fig:Pf_compare}. However, for all cases considered with $T > 1$ eV, the $g(r)$ and self diffusion coefficient are adequately described by a RPP. With the force-matched-validated NPA interaction, precomputing the interaction allows for much larger pair-potential simulations.

As fast analytic expressions for transport coefficients are needed for hydrodynamic modeling, we compared our self diffusion results from all models to the Stanton-Murillo model for both C and V. In both cases, the Stanton-Murillo model was consistent with the TFY model (on which it is based) and both have agreement with force-matched-based results. The Stanton-Murillo error relative to force-matched is $<65\%$ below $T = 10$ eV for V and $<25\%$ below $T = 5$ eV for C, adding confidence to the use of this model in hydrodynamics models above that temperature. For experiments that are rapidly heated above a few eV, little time is spent where the errors are large; because the transport coefficients are numerically very small during this transient heating, negligible transport can occur during that time. For example, note that the V diffusion coefficient varies by a factor of about $30$ in the range $T = 0.5$  to $100$ eV. Conversely, for experiments that dwell at lower temperatures, we provide a RPP-based correction factor to the Stanton-Murillo model with an error of less than 1\% for C at $T = 0.5$ eV and $6\%$ for V at $T = 0.5$ eV.

Our results suggest several new avenues of investigation. From a data science perspective, larger collections of systematically-obtained simulation results would aid in better defining accuracy boundaries. In particular, more elements that produce more material types should be studied. For mixtures, $N$-body potentials could be explored; here, we cast all of the pair potentials as heteronuclear. Additionally, our conclusions are based on studies of $g(r)$ and self-diffusion, which could be extended to include other properties such as viscosities and interdiffusion in mixtures. Finally, as very large scale simulations become more common, spatially heterogeneous plasmas can be modeled; much less is known about potentials of any order in such environments, although recent work has explored non-spherical potentials \cite{stanton2018multiscale}.

\begin{acknowledgments}
The authors would like to thank Jeffrey Haack, Liam Stanton, Patrick Knapp, and Stephanie Hansen for insightful discussions, and Josh Townsend who provided VASP post-processing tools. LAMMPS can be accessed at http://lammps.sandia.gov.  Sandia National Laboratories is a multimission laboratory managed and operated by National Technology \& Engineering Solutions of Sandia, LLC, a wholly owned subsidiary of Honeywell International Inc., for the U.S. Department of Energy's National Nuclear Security Administration under contract DE-NA0003525. This paper describes objective technical results and analysis. Any subjective views or opinions that might be expressed in the paper do not necessarily represent the views of the U.S. Department of Energy or the United States Government SAND2020-13783 O. %
\end{acknowledgments}

\begin{appendix}

\section{NPA Details and Extensions}

In this appendix, we provide a brief background on NPA calculations, and specific extensions needed for the argon, iron and vanadium cases at low temperatures. While we believe the results are formally correct for these cases, we address the nature of the questions being asked and the extensions that are needed.

\subsection{NPA Formulation}
The NPA model applies primarily to warm-dense fluids with spherical symmetry, although the NPA
method can be applied equally well to crystals~\cite{Dagens75, PeBe}. Importantly, there is no unique NPA model ~\cite{StaSauDaligHam14, murillo2013partial, Rosznayi08, SternZbar07}; here, we describe a specific set of choices ~\cite{cdw-N-rep19,eos95,PeBe,DWP1} based around a formal statement of the theory ~\cite{DWP1}. A key difference between many average-atom models and the NPA is that the free electrons are not confined to the Wigner-Seitz sphere, but
 move in all of space as approximated by a very large correlation
 sphere, of radius $R_c$ which is ten to twenty times the Wigner-Seitz
 radius~\cite{murillo2013partial}.

Our NPA model begins with the variational property of the grand potential
 $\Omega[n_e,n_i])$ as a functional of  the  one-body densities
  $n(r)$ for electrons, and $\rho(r)$ for ions. Only a single nucleus
of the material is used and taken as the center of the
 coordinate system. The other ions (``field ions'') are replaced by
 their one-body density distribution $\rho(r)$: DFT asserts that
 the physics is solely given by the one-body distribution; i.e.,
 we do not need two-body, three-body, and such information as they
 get included via exchange-correlation (XC)-functionals. Note that this 
 formulation differs from $N$-center codes \cite{VASP, ABINIT} like the VASP or ABINIT.
Moreover, there can be other differences; for example, we have used the finite-$T$ electron XC-functional by Perrot and Dharma-wardana~\cite{PDWXC}, while the PBE 
implemented on VASP is a $T=0$ XC-functional. The finite-$T$ functional
used is in good agreement with quantum Monte-Carlo XC-data~\cite{cdw-N-rep19}
in the density and temperature regimes of interest.
 
The artifice of using a nucleus at the origin converts the one body ion
 density $\rho(r)$ and the electron density $n(r)$ into effective two body
 densities in the sense that
\begin{equation}
n_i(r)=\bar{n}_ig_{ii}(r),\; n_e(r)=\bar{n}_eg_{ei}(r)
\end{equation}
The origin need not be at rest; however, most ions are heavy enough that
the Born-Oppenheimer approximation is applicable.
Here $\bar{n}_i,\bar{n}_e$ are the mean ion density and the mean
 free electron density, respectively. Bound electrons are assumed to be firmly associated with each ionic nucleus and contained in their ``ion cores'' of radius $r_c$ such that
\begin{equation}
\label{core.eqn}
 r_c<a_i. 
\end{equation}
In some cases, e.g., some transition metals, and for continuum resonances etc.,
 this condition for a compact core may not be met, and additional
 steps are needed. We assume a compact core as a working hypothesis.
 The DFT variational equations used here are:
\begin{equation}
\label{KS-basic.eqn}
\frac{\delta \Omega[n_e,n_i]}{\delta n_e}=0;\;\;\frac{\delta \Omega[n_e,n_i]}{\delta n_i}=0.
\end{equation}
These directly leads to two coupled Kohn-Sham equations where the unknown quantities are the XC-functional for the electrons, and the ion-correlation
functional for the ions~\cite{ilciacco93}. If the Born-Oppenheimer
approximation is imposed, the ion-electron XC-functional may also be
neglected. Approximations arise in modeling these functionals and
in decoupling the two Kohn-Sham  equations~\cite{eos95,CPP} to some
extent, for easier numerical work.  The first equation gives the usual
Kohn-Sham equation for electrons moving in the external potential of
 the ions. This is the only DFT equation used in $N$-center codes in which
 ions define a periodic structure evolved by MD, followed by a Kohn-Sham solution at
 each step. In contrast, NPA employs the one-body ion density $n_i(r)$; it was shown in ~\cite{DWP1} that the ionic DFT equation can be identified
 as a Boltzmann-like distribution of field ions around the
 central ion, distributed according to the `potential of mean force'
 well known in the theory of fluids. In such a formulation, the ion-ion correlation functional $F_{xc}^{ii}$ was identified to be the sum of hypernetted-chain diagrams plus the bridge diagrams as an exact result formally, although the bridge diagrams cannot be evaluated exactly.

The mean electron density  $\bar{n}_e$ can also be specified as the number of
 free electrons per ion, viz., the mean ionization state $\langle Z \rangle$. Although the material density $\bar{n}_i$ is specified, the mean free electron
 density $\bar{n}_e$ is unknown at any given temperature, as it depends
 on the ionization balance which is controlled by the free energy
 minimization~\ref{KS-basic.eqn}. Hence, a trial value for $\bar{n}_e$ (i.e, equivalently, a trial value for $\langle Z \rangle$) is assumed and the thermodynamically consistent $n_i(r)$ is
determined. This is repeated until the target mean ion density $\bar{n}_i$
 is obtained.  

This means that the Kohn-Sham equation has to be solved for a single
 electron moving in the field of the central ion;  its ion distribution
 $\bar{\rho}g(r)$ is modified at each iteration with modification
of the  trial $\bar{Z}$ until the
 target material density is found. However, it was noticed very early~\cite{PeBe, eos95} 
 that the Kohn-Sham solution was quite insensitive to the details of the $g(r)$ and hence a
 simplification was possible. The simplification was to replace the trial
 $g(r)$ at the trial $\bar{Z}$ by a cavity-like distribution:
\begin{equation}
\label{cavity-gr.eq}
g_{cav}(r)=0, r\le a_i, \; g_{cav}(r)=1, r>a_i
\end{equation}
Here the $a_i$ is the trial value of $a_i$,
 based on the trial $\bar{n}_e$. Hence, adjusting the $g_{cav}(r)$ at each
 iteration requires only adjusting the trail $a_i$ to achieve
 self-consistency. The self-consistency in the ion distribution is
 rigidly controlled by the Friedel sum rule for the phase shifts of
the Kohn-Sham-electrons~\cite{DWP1}. This ensures that $\bar{\rho}=\bar{n}/Z$.
Thus, a valuable result of the calculation using the cavity model
 of $g(r)$ is the self-consistent value of the mean ionization state $Z$,
 which is both an atomic quantity and a thermodynamic quantity.

Here we note crucial simplifications used in implementing
the NPA. Given that the electron distribution $n(r)$ obtained self 
consistently can be written as a bound-electron term
and a free-electron term because of the condition specified by
Eq.~\ref{core.eqn}, we have:
\begin{equation}
\label{pileup.eqn}
n_e(r)=c(r)+n^f(r), \; \Delta n_e(r)=n^f(r)+\bar{n}_e
\end{equation} 
The core-electron density (made up of ``bound electrons'') is denoted by
$c(r)$. The free electron density $n_f(r)$ is the response of an electron fluid
 containing a cavity that mimics $n_i(r)$. It contributes
 to the potential on the electrons. The response of a uniform electron gas
to the central ion can be obtained by subtracting the
effect of the cavity using the known static
 interacting linear response function  $\chi(k,\bar{n}_e,T)$ of the
 electron fluid. That is, from  now on we take it that the charge
 density $n_f(r)$ and the charge pile up  $\Delta n_f(r)$ are both
 corrected for the presence of the cavity, but we use the same
 symbols.

\subsection{Pair-potentials for NPA Mixtures: Ionic Contributions}
\label{pots-npa.sec}

With the basic NPA formulated, we turn to the construction of pair potentials, with a focus on the more challenging cases we consider in the main text. It is very common to treat WDM and liquid metals through purely ionic interactions, which are adequately evaluated in second-order perturbation theory unless the free electron density and the temperature ($T/E_F$) are very low. Such interactions, which generalizes (\ref{sopt}) to mixtures, so are written in $k$-space as
\begin{eqnarray}
\label{ppot-f.eqn}
U_{ab}(k)&=&Z_aZ_bV_k +\\
         & & \{U^{ei}_a(k)\Delta n^f_b(k)+
U^{ei}_b(k)\Delta n^f_a(k)\}/2 \\
V_k&=&4\pi/k^2. 
\end{eqnarray}
In the NPA theory for mixtures, $Z_a$ and $Z_b$ are integers, while
in the simple (average-atom) NPA, the $\bar{Z_s}=\Sigma_s x_sZ_s$ is used. 
The density fluctuations use the linear-response property of the pseudopotential
\begin{equation}
\Delta n^f(k)=U^{ei}(k)\chi(k).
\end{equation}
Here, since  $n^f(k)$ has been calculated via Kohn-Sham, it has all the
 non-linear effects included by the construction of $U^{ei}(k)$.
 The extent of the validity of such a quasi-linear pseudopotential is discussed in Ref.~\cite{Utah12}. Unlike in the average atom NPA, the $U^{ei}$ used in mixture-theory
is the pseudoptential of the ion with the appropriate 
{\it integer ionization}. The interacting electron gas response function used in these calculations includes a local-field factor chosen to satisfy the finite temperature
electron-gas compressibility sum rule, and is given explicitly by
\begin{eqnarray}
\label{resp.eq}
\chi(k,T_e)&=&\frac{\chi_0(k,T_e)}{1-V_k(1-G_k)\chi_0(k,T_e)},\\
\label{lfc.eq}
G_k &=& (1-\kappa_0/\kappa)(k/k_\text{TF});\quad V_k =4\pi/k^2,\\
\label{ktf.eq}
k_{\text{TF}}&=&\{4/(\pi \alpha r_s)\}^{1/2};\quad \alpha=(4/9\pi)^{1/3}.
\end{eqnarray}
Here $\chi_0$ is the finite-$T$ Lindhard function, $V_k$ is the bare Coulomb
potential, and $G_k$ is a local-field correction (LFC). The finite-$T$
compressibility sum rule for electrons is satisfied since $\kappa_0$ and
$\kappa$ are the non-interacting and interacting electron compressibilities
respectively, with  $\kappa$ matched to the $F_{xc}(T)$ used in the Kohn-Sham
calculation. In Eq.~\ref{ktf.eq}, $k_\text{TF}$ appearing in the LFC is the
Thomas-Fermi wave vector. We use a $G_k$ evaluated at $k\to 0$ for all $k$
instead of the more general $k$-dependent form (e.g., Eq.~50  in
Ref.~\cite{PDWXC}) since the $k$-dispersion in $G_k$ has negligible
effect for the WDMs of this study.

\subsection{Pair-potentials for NPA Mixtures: Core Interactions}
\label{core-pots-npa.sec}

We now consider extensions to this formulation that includes the effect of core electrons
(see Appendix B of Ref.~\cite{eos95}). Core-core interactions are important for atoms
like argon, sodium, potassium, gold, etc., with large cores and 
zero or low $\langle Z \rangle$. Here we use a simplified approach in discussing the
case of low-$T$ argon rather than the more exact approach given in
 Ref.~\cite{eos95}. The total pair-interaction $\psi_{ab}$
 is of the form:
\begin{eqnarray}
\label{ppot-F.eqn}
\psi_{ab}&=&U(c_a,c_b)+\{U(c_a,f_b)+U(f_a,c_b)\}+U(f_a,f_b)\nonumber\\
         &=&U^{cc}(r)+U^{cf}(r)+U^{ff}(r)
\end{eqnarray}
The first term, $U^c=U(c_a,c_b)$ is the interaction between the
 two atomic cores, and  it is the only term found in a neutral gas
 of pure argon atoms. 
 The second term is the interaction of the core electrons of the neutral atom $a$ with the pseudo-potential of the
 ion $b$ with integer charge $Z_b$,
while the indices are interchanged in the third term.  
The last term is the  corresponding DFT corrections
to the exchange-correlation energy and to the interacting kinetic energy.
How to evaluate these in the NPA at any temperature and density
is given in the Appendix B of Ref.~\cite{eos95} to the same level
of  approximation as Eq.~\ref{ppot-f.eqn}, i.e., to second-order in
perturbation theory in the screened interactions. It is found 
that such evaluations for argon give results close to parametrized
potentials similar to the Lennard-Jones(LJ) or more sophisticated 
potentials, but inclusive of a screening correction. Thus, at the
LJ-level of approximation, $U^c(k)$ for two neutrals immersed in
the electron gas is approximately given by
 $U^{LJ}(k)\{1+V_k\chi(k)\}$. For two charged ions, a correction
factor of $\{(Z_n-Z_{int})/Z_n\}^2$, where $Z_n$ is the nuclear charge, and
$Z_{int}$ is the integer-ionization of the ion is included because the
ion cores have less electrons than the neutral cores.

\subsection{Argon in low-$T$ WDM states}

The NPA calculation at density $\rho \simeq$1.4 g/cm$^3$ and temperature $T = 2$eV
 yields a mean charge of $\langle Z \rangle \simeq 0.3$; however, this implies a mixture where
30\% of the argon atoms are in the Ar$^+$ state, while 70\% of the atoms
are neutral atoms. We ignore the Ar$^{2+}$ ionization state as the second ionization energy is about $27$ eV (ignoring plasma corrections). So, while the NPA  tries to
assign a mean ionization state, this argon system is better treated as a two component mixture with $x_a, x_b,$ where $a=Ar, b=Ar^+$. 

Thus, we need the three pair-potentials $U_{ab}$, i.e., for  Ar$-$Ar, Ar$-$Ar$^+$,
and Ar$^+-$Ar$^+$  interactions. Here the atomic species are immersed
 in the electron gas  resulting from the ionization. These 
pair-interactions can be rigorously calculated from first-principles
using the atomic and ionic electron densities obtained in the NPA,
as discussed in Appendix B of reference~\cite{eos95}. Here we follow
a more simplified calculation of the pair potentials exploiting
known models for argon, suitable for this mixture with 70\% neutral
argon.

The Ar$^+$ ion
interacts with the neutral Ar atom by polarizing the core-electron
distribution. This distortion is usually described in terms of
the dipole polarizability $\alpha$ and quadrupole polarizability
$\beta$  of the Ar atom, {\it viz.} 
\begin{eqnarray}
\label{ci-pot.eqn}
U^{cf}(r)&=&U(c_a,f_b)+U(c_b,f_a), \\
   U(c_a,f_b)(r)&=& C_4/r^4 +C_6/r^6+...,\\
              C_4&=&\alpha(Z_{int})^2/2,\; C_6=\beta(Z_{int})^2/2.
\end{eqnarray}
This interaction will be screened by the polarization processes of the
background electron gas. Instead of using the truncated
multipole expansion, an improved calculation may be done as
in Appendix B of Ref.~\cite{eos95}. The ion-atom contribution from
Eq.~\ref{ci-pot.eqn} contains only the last term, as the atom $b$ is
neutral and $U^{ei}_b(k)=0$. In the regime $T=2$ eV for Ar, we have simply
approximated the core-core interaction of the atom-ion interaction by the
mean value of atom-atom and ion-ion core-core interaction.

\begin{figure}[t]
\begin{center}
\includegraphics[width=\columnwidth]{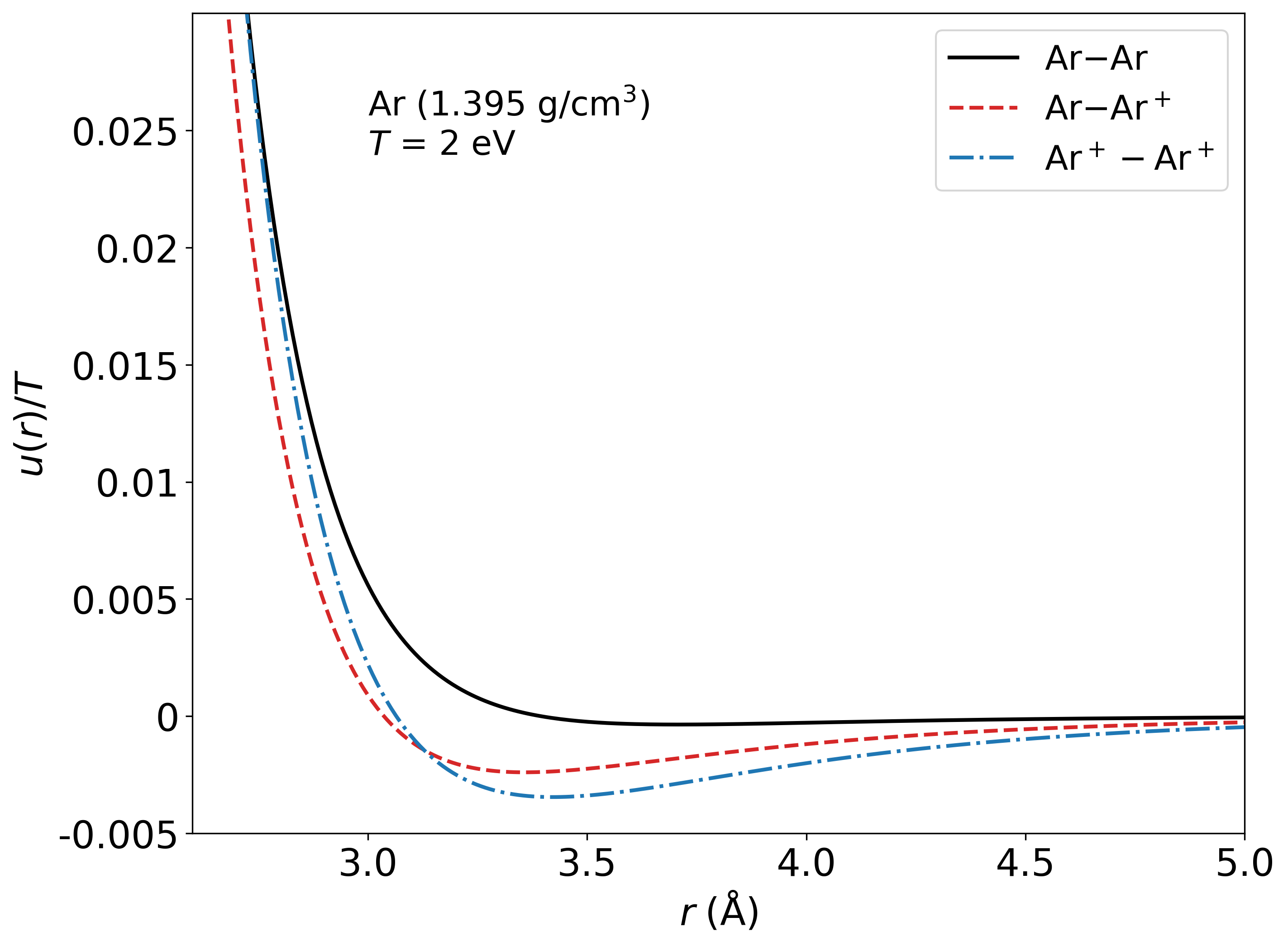}
\label{argonpots.fig}
\caption{(Online color)The pair potentials among the Ar and Ar$^+$
species of the two component mixture of  argon at 2 eV}   
\end{center}
\end{figure}

In Figure \ref{argonpots.fig} we display the pair potentials for
Ar$-$Ar, Ar$^+-$Ar, and Ar$^+-$Ar$^+$ at $T=2$ eV at a density of
1.395 g/cm$^3$. The compositions 0.7 for Ar, and 0.3 for Ar$^+$ are obtained from
the simple NPA calculation. We have not attempted to optimize these
composition fractions using steps indicated in Ref.~\cite{eos95}.

\subsection{ Iron and Vanadium in low-$T$ WDM states}
The NPA uses an ``isotropic'' atomic model even for iron, vanadium and other
transition metals, and ignores the $s-d$ hybridization energy $E_{hyb}$ 
that re-arranges the electron distribution among the $d$ and $s$ electron
 states near the Fermi energy. The need for $s-d$ hybridization is best seen
 by looking at the low temperature band  structure of such a transition
 metal. The unhybridized bands of an ``isotropic model'' for vanadium are
 such that the free-electron like band crosses the $d$-bands, and the $s-d$
 interaction redistributes the electrons. Furthermore, instead of the $s$-wave
local pseudopotential (as employed here) an angular-momentum dependent
form is more appropriate at low-$T$.  Hence the calculation of the
 mean ionization has to be  accordingly modified. However, the simpler
 picture re-emerges when the temperature exceeds the $s-d$ hybridization
 energy.

Furthermore, charge polarization fluctuations of the $d$-electrons can
 couple with those of the electron gas (e.g., as discussed  by
 Maggs and Ashcroft \cite{Maggs87}). Such effects, as well as ``on-site
 Hubbard U'' effects
need to be included in the electron XC-functionals to properly treat
 transition metals. The usual XC-functionals made available with standard
 codes do not include these effects. However, at sufficiently high
temperatures the ionization becomes high enough to screen these effects
and the theory simplifies.
The pair-potentials for iron provided by the isotropic model with no
$s-d$ hybridization and other $d$-band effects lead to cluster formation
at low-$T$. Here, unlike in liquid carbon or silicon, the bonding is not
transient. Hence low-$T$ MD simulations will show no movement of the ions
and no diffusion. %
In the case of silicon and carbon, both of which are known to show varying degrees of transient bonding behavior, NPA was shown to produce an accurate description of the high density liquid phase, with structure factors and PDFS obtained from the NPA-HNC procedure agreeing closely~\cite{cdwSi20}  with those from 218-atom simulations with DFT.

\end{appendix}

\bibliography{force_matching}%

\end{document}